# Pore-scale study of dissolution-induced changes in hydrologic properties of rocks with binary minerals

Li Chen (lichennht@gmail.com)

a: Key Laboratory of Thermo-Fluid Science and Engineering of MOE, School of Energy and Power Engineering, Xi'an Jiaotong University, Xi'an, Shaanxi 710049, China b: Earth and Environmental Sciences Division, Los Alamos National Laboratory, Los Alamos 87545, New Mexico, USA

Qinjun Kang (qkang@lanl.gov)

b: Earth and Environmental Sciences Division, Los Alamos National Laboratory, Los Alamos 87545, New Mexico, USA

Hari S. Viswanathan

b: Earth and Environmental Sciences Division, Los Alamos National Laboratory, Los Alamos 87545, New Mexico, USA

Wen-Quan Tao

a: Key Laboratory of Thermo-Fluid Science and Engineering of MOE, School of Energy and Power Engineering, Xi'an Jiaotong University, Xi'an, Shaanxi 710049, China

**Corresponding author**

Qinjun Kang, Computational Earth Group, Earth and Environmental Sciences Division, Los Alamos National Laboratory, Los Alamos 87545, New Mexico, USA (qkang@lanl.gov)

**Abstract**

A pore-scale numerical model for reactive transport processes based on the Lattice Boltzmann method is used to study the dissolution-induced changes in hydrologic properties of a fractured medium and a porous medium. The solid phase of both media consists of two minerals, and a structure reconstruction method called quartet structure generation set is employed to generate the distributions of both minerals. Emphasis is put on the effects of undissolved minerals on the changes of permeability and porosity under different Peclet and Damkohler numbers. The simulation results show porous layers formed by the undissolved mineral remain behind the dissolution reaction front. Due to the large flow resistance in these porous layers, the permeability increases very slowly or even remains at a small value although the porosity increases by a large amount. Besides, due to the heterogeneous characteristic of the dissolution, the chemical, mechanical and hydraulic apertures are very different from each other. Further, simulations in complex porous structures demonstrate that the existence of the porous layers of the nonreactive mineral suppresses the wormholing phenomena observed in the dissolution of mono-mineralic rocks.

**Keyword:** reactive transport; dissolution; permeability; porous media; fracture; lattice Boltzmann method

# 1. Introduction

Reactive transport phenomena involving dissolution-precipitation are pervasive in a variety of scientific, industrial and engineering processes. Typical examples include self-assembled patterns such as Liesegang rings or bands [*Chen et al.*, 2012a], biofilm growth in aqueous environment [*Pintelon et al.*, 2012], environmental contaminant transport [*Cubillas et al.*, 2005], extracting hydrocarbon from unconventional resources such as low permeability shales [*Yao et al.*, 2013], acid injection for enhanced oil recovery [*Maheshwari et al.*, 2013], and geologic sequestration of carbon dioxide ($CO_2$) in deep saline aquifers [*Stockmann et al.*, 2011]. The dissolution-precipitation can lead to modification of the solid phase structures. Such modification would remarkably change the hydrologic properties including structural parameters (porosity, tortuosity and surface area, etc) and transport properties (permeability, effective diffusivity, etc) of the porous media [*Luquot and Gouze*, 2009; *Luquot et al.*, 2014; *Noiriel et al.*, 2009]. The evolutions of these properties are controlled by several factors including the physicochemical conditions of the pore fluid, the porous geometries and the composition of the solid phase [*Luquot and Gouze*, 2009]. A better understanding and characterization of the evolution of these properties is critical for applications such as hydraulic fracturing and geologic $CO_2$ sequestration.

Processes that cause the changes of hydrologic properties often occur at the pore scale. However, hydrologic properties are defined at the continuum scale. This mismatch in length scales makes it difficult to correctly understand and describe the changes of hydrologic properties at the continuum scale. Pore-scale studies are necessary to better characterize these changes, because detailed information from pore-scale modeling can be used to derive hydrologic properties and constitutive relationships; and time-dependent information can be used to characterize their changes with time [*Békri et al.*, 1995; 1997; *Chen et al.*, 2014a; *Chen et al.*, 2013b; *Huber et al.*, 2013; *Kang et al.*, 2003; *Kang et al.*, 2006; *Kang et al.*, 2002b; *Kang et al.*, 2010; *Kang et al.*, 2014; *Li et al.*, 2008; *Luo et al.*, 2012; *Molins et al.*, 2012; *Alexandre M. Tartakovsky et al.*, 2007; *A. M. Tartakovsky et al.*, 2007; *Tartakovsky et al.*, 2008; *Yoon et al.*, 2012].

Variations of hydrologic properties during dissolution processes strongly depend on physicochemical processes. The relative strengths of convection, diffusion and reaction can be well represented by two important dimensionless numbers in reactive transport process, namely the Peclet number $Pe$ (relative strength of fluid flow to diffusion) and Damkohler number $Da$ (the relative strength of chemical reaction to diffusion). Many pore-scale studies have been performed to study the effects of $Pe$ and $Da$ on dissolution processes and the variations of hydraulic properties. Bekri et al. [1995, 1997] first performed pore-scale simulations of dissolution processes by combining finite different and cellular automaton (CA) methods for reactive transport and solid structure update, respectively. Four distinct dissolution patterns were identified based on the $Pe$ and $Da$: (a) for small $Pe$ and small $Da$, dissolution is uniform in the entire domain; (b) for large $Da$ and large $Pe$, dissolution occurs along main channel parallel to the flow direction (wormhole phenomenon); (c) for $Da$ and $Pe$ of order one, dissolution occurs isotropically around the central cavity and symmetrically along the flow direction; and (d) for large $Da$ and small $Pe$, dissolution occurs around the central cavity and then along the main channels [*Békri et al.*, 1995]. Kang et al. [2002b, 2003] proposed a pore-scale model based on the lattice Boltzmann method (LBM) and CA to explore the dissolution processes. Their studies substantiated the dissolution patterns under different $Pe$ and $Da$ numbers [*Kang et al.*, 2003]. They also demonstrated that an optimal injection rate exists at which the wormhole is formed as well as the amount of reactant for breakthrough is minimized, and the optimal injection rate is affected by the *Pe* and *Da* [*Kang et al.*, 2002b].

Based on the above pore-scale studies, it can be concluded that, while porosity can be directly related to the dissolution amount of the solid phase, the determination of the permeability is more complex, as it not only depends on the dissolution amount but also is affected by the dissolution patterns. The description of the dissolution patterns is very complex depending on the physicochemical conditions. Consequently, for the same porosity the permeability varies under different *Pe* and *Da*, as shown by the simulation results in the literature [*Békri et al.*, 1995; *Huber et al.*, 2013; *Kang et al.*, 2003; *Sadhukhan et al.*, 2012], indicating that the permeability is not generally a simple function of the porosity. Currently, the Kozeny-Carman (KC) equation is the most widely used empirical law relating permeability to porosity [*Carman*, 1937]. However, as it is based on the assumption of uniform changes of the solid phase, it cannot account for the variation of the permeability on the porosity under different dissolution patterns, especially when strong heterogeneity of the solid structure modification takes place, such as wormhole phenomenon [*Daccord et al.*, 1993].

While much progress has been made to understand the effects of the physicochemical processes on hydrologic properties, the composition of the solid phase, an important factor which also plays an important role [*Noiriel et al.*, 2007; *Noiriel et al.*, 2013], has received little attention. Experimentally, not until recently have the effects of the undissolved minerals on the dissolution patterns as well as on the variations of hydrological parameters been emphasized [*Noiriel et al.*, 2007; *Noiriel et al.*, 2013]. Natural rocks usually consist of multiple minerals. In the experiment of Noiriel et al. [2007], several minerals were detected including calcite, clay, quartz, dolomite and pyrite in the argillaceous limestone rock sample. Different minerals present different dissolution rate. For example, in the experiment of Noiriel et al. [2007], the dissolution rate of clay is about $10^6$ magnitude lower than that of carbonate minerals. Significant difference on the dissolution rate will greatly affect the geometrical evolutions of the rocks which in turn affect the reactive transport processes and hydraulic properties [*Noiriel et al.*, 2007; *Noiriel et al.*, 2013]. A microporous coating formed by the undissolved minerals was identified experimentally, which greatly hinders the mass transport and increases the flow resistance. The movement and reorganization of the undissolved minerals made the reactive transport processes more complicated. The dissolution processes and the permeability were greatly affected by the undissolved minerals [*Noiriel et al.*, 2007]. However, to our best knowledge, all of the published pore-scale simulations focusing on changes of hydrological properties have been performed with monomineralic rock.

The present work is mainly motivated by the experiments of Noiriel et al. [2007] and Noiriel et al. [2013], where effects of the undissolved minerals were studied. The pore-scale numerical model adopted is developed by Kang and co-authors in a series of publications [*Chen et al.*, 2013b; *Kang et al.*, 2003; *Kang et al.*, 2006; *Kang et al.*, 2002b]. The remaining part of the present study is arranged as follows. In Section 2, the quartet structure generation set (QSGS), a structure reconstruction method, is adopted to generate the distributions of different minerals in rocks. Then in Section 3, the physicochemical model proposed is introduced and the pore-scale model employed is presented. In Section 4, dissolution process in a fractured porous medium is studied and the effects of the undissolved mineral on the reactive transport processes and the variation of porosity, permeability and fracture apertures are discussed in detail. Dissolution in a complex porous medium is also explored in Section 5, where the effects of the undissolved mineral on the formation of wormhole phenomenon are discussed. Finally, some conclusions of the present study are drawn in Section 6.

## 2. Structure generation of rocks with binary minerals

A detailed description of the distributions and morphology of the multiple minerals in rocks is required for accurate predictions of the related reactive transport and dissolution processes. This can be directly obtained from experimental techniques with high resolution such as the Scanning electron microscopy (SEM). Reconstruction algorithms provide an alternative method for reproducing such distributions. In the present study, the quartet structure generation set (QSGS) developed by Wang et al. [2007] is adopted for reconstructing rocks with multiple minerals. The QSGS has been demonstrated to be capable of generating morphological features closely resembling the forming processes of many real porous media [*Wang et al.*, 2007].

Without loss of generality, rocks with binary minerals are considered, which can reduce the structure complexity, but still keeps the elemental characteristic for investigating the effects of coexistence of multiple minerals with different dissolution rates. In the rocks, the main mineral is α and the discrete mineral is β. Without loss of generality, the main mineral α is determined as non-growing phase and the discrete mineral β is growing phase in the QSGS. The reconstruction starts with rocks in certain regions of the computational domain. The distributions of rocks are either ideally specified (Section 4) or are rather complex as in real porous media (Section 5), with the former one for clearly understanding the underlying mechanisms and the later one taking into account the realistic porous structures. The procedures of generating α and β distributions in the rocks using the QSGS are briefly described as follows, and for more details one can refer to Wang et al. [2007].

(1) Randomly locate the cores of β in the rock based on a distribution probability, $c_d$, which is smaller than the volume fraction of β in the rock, $\varepsilon_\beta$. A random number within (0,1) will be assigned to each node in the rock, and if the random number is no greater than $c_d$ the node is selected as a core.

(2) Grow an existing β node to its neighboring nodes based on another probability $P_i$ with $i$ representing the direction. In the present study only the nearest neighboring nodes are considered, namely four in the main direction (1,2,3,4) and the other four in the diagonal directions (5,6,7,8), as shown in Fig. 1. For each direction, a new random number is generated, and if it is no greater than $P_i$ the neighboring node in $i$ direction becomes β.

(3) Repeat (1) and (2) until the volume fraction of β, namely $\varepsilon_\beta$, is achieved.

(4) The remaining nodes not occupied by β in the rocks are the main mineral α.

There are three parameters ($c_d$, $P_i$, $\varepsilon_\beta$) controlling the generated morphologies of α and β. $c_d$ denotes the number density of the growing cores of β in the rocks. A higher value of $c_d$ leads to more cores in the system, resulting in more particles of β, while a lower value leads to less cores generating bigger β agglomerates and thus increases the computation fluctuation [*Wang et al.*, 2007]. $P_i$ is the probability of the growth of an existing β grid towards its neighboring grid in the $i$ direction. In order to generate isotropic structures, $P_i$ of the four main directions (1,2,3,4) should be the same, so is that for the four diagonal directions (5,6,7,8) [*Wang et al.*, 2007]. Antistrophic structures can be obtained by simply setting different values of $P_i$ in different directions.

Fig. 2 shows the distributions of α and β in the rock generated by the QSGS method, where $c_d$ and $\varepsilon_\beta$ are the same for all the images. The dark color represents mineral α, the gray stands for the β and the white part denotes the void space. In Fig.2(a), $P_{(1,2,3,4)}$ are the same and the ratio of $P_{(1,2,3,4)}$ to $P_{(5,6,7,8)}$ is 4. The random distributions of β in the rock and the isotropic structures of β can be clearly observed in Fig. 2(a). The structure features of the undissolved mineral plays an important role on the hydraulic and transport properties during the dissolution of rocks[*Noiriel et al.*, 2007]. By choosing different directional growth probabilities in different directions, antistrophic structures of the β particle can be generated as shown in Fig. 2(b) and 2(c). In Fig. 2(b), $P_{(1,3)}$, which are in the *x* direction, are set as 0.0004, while the value of $P_i$ in remaining six directions are zero. It can be seen that β particles are elongated along horizontal direction. In Fig. 2(c), $P_{(2,4)}$ are set as 0.0004 while the value of $P_i$ in remaining directions are zero, leading to vertically elongated β particles. As will be discussed in Section 4, different structure features of the β particle significantly affect permeability.

## 3. Physicochemical model and numerical method

### 3.1 Physicochemical model

Our study focuses on effects of the undissolved minerals on the reactive transport processes and changes of structural and hydraulic parameters during dissolution. Generally, rocks in natural systems consist of multiple minerals [*Noiriel et al.*, 2007]. Different minerals have different dissolution rates. Significant difference of the dissolution rate will greatly affect the geometrical evolutions of the rocks which in turn affect the reactive transport processes and hydraulic and transport properties [*Noiriel et al.*, 2007; *Noiriel et al.*, 2013]. In most of our simulations in this study, the main mineral α is the dissolving mineral and the discrete mineral β will not dissolve. Following our previous work, the dissolution reactions at the dissolving mineral-fluid interface are generalized as [*Kang et al.*, 2002b]

$$R_{(aq)} + \alpha_{(s)} \leftrightarrows P_{(aq)} \tag{1}$$

where $R_{(aq)}$ is the reactant and $P_{(aq)}$ the product. Subscript "aq" and "s" stand for aqueous phase and solid phase, respectively. Eq. (1) can be considered as a simplified form of several realistic dissolution reactions in natural systems and experiments. For example, during $CO_2$ sequestration in deep saline aquifers, the progressive dissolution of $CO_2$ in the fluid leads to a decrease of pH. As a result, dissolution of the carbonates takes place and the dissolution reaction can be described by Eq. (1) where $R_{(aq)}$ is $H^+$, α is $CaCO_3$, $P_{(aq)}$ is $Ca^{2+}$ and β is clay [*Noiriel et al.*, 2007]. The technique of acid injection for enhancing oil recovery is another typical example. Acid such as hydrochloric acid HCl is injected into the rocks and dissolution of the minerals occurs, which increases the permeability of the oil reservoirs and thus facilitates the oil exploration [*Maheshwari et al.*, 2013]. The dissolution reaction between acid and the rocks can also be simply described using Eq. (1).

Based on the assumption that the dissolution occurs by a first-order chemical reaction, the dissolution rate $r_\alpha$ [mol m$^{-2}$s$^{-1}$] is given by [*Kang et al.*, 2002b]

$$r_\alpha = k_r (C_{R_{(aq)}} - C_{P_{(aq)}} / K_{eq}) \tag{2}$$

where $k_r$ [m s$^{-1}$] is the effective forward reaction rate constant, $K_{eq}$ is the effective equilibrium constant and $C$ is the concentration at the reactive surface.

The general physicochemical processes can be described as follows. Initially, rocks are in equilibrium with R$_{(aq)}$ and P$_{(aq)}$ in the domain. Then reactant R$_{(aq)}$ is injected into the domain, resulting in disequilibrium of the system. Dissolution of α takes place, consuming R$_{(aq)}$ and generating P$_{(aq)}$. The undissolved mineral remains in the system. In the present study, we assume the solute transport has negligible effects on the fluid flow, which is reasonable as the solute concentration is sufficient low in lots of realistic scenarios. The governing equations for fluid flow and solute transport are as follows.

$$\frac{\partial \rho}{\partial t} + \nabla \cdot (\rho \mathbf{u}) = 0 \tag{3a}$$

$$\rho \frac{\partial \mathbf{u}}{\partial t} + \rho (\mathbf{u} \cdot \nabla) \mathbf{u} = -\nabla p + \nabla \cdot (\nabla \mu \mathbf{u}) \tag{3b}$$

$$\frac{\partial C_{R_{(aq)}}}{\partial t} + (\mathbf{u} \cdot \nabla) C_{R_{(aq)}} = D_{R_{(aq)}} \Delta C_{R_{(aq)}} \tag{3c}$$

$$\frac{\partial C_{P_{(aq)}}}{\partial t} + (\mathbf{u} \cdot \nabla) C_{P_{(aq)}} = D_{P_{(aq)}} \Delta C_{P_{(aq)}} \tag{3d}$$

where $t$ is time, $\rho$ fluid density, $p$ pressure, $\mu$ dynamic viscosity, **u** the velocity and $D$ the diffusivity. Eqs. 3(a)-(b) are the Navier-Stokes (NS) Equation and Eqs. 3(c)-(d) are the advection-diffusion equations for the concentration of $R_{(aq)}$ and $P_{(aq)}$, respectively.

At the fluid-mineral α interface, dissolution reaction takes place which consumes $R_{(aq)}$ and releases $P_{(aq)}$ as described by Eq. (1). With first-order reaction kinetics model expressed by Eq. (2), the boundary conditions corresponding to the dissolution reaction for $R_{(aq)}$ and $P_{(aq)}$ are as follows

$$D_{R_{(aq)}} \frac{\partial C_{R_{(aq)}}}{\partial n} = k_r (C_{R_{(aq)}} - C_{P_{(aq)}} / K_{eq}) \tag{4a}$$

$$D_{P_{(aq)}} \frac{\partial C_{P_{(aq)}}}{\partial n} = -k_r (C_{R_{(aq)}} - C_{P_{(aq)}} / K_{eq}) \tag{4b}$$

where $n$ is the direction normal to the reactive surface pointing to the fluid. $\partial C / \partial n$ is the concentration gradient at the fluid-mineral α interface.

The dissolution reaction causes the volume loss of the mineral α. The volume change of a solid node under dissolution reaction is calculated by

$$\frac{\partial V_{\alpha_{(s)}}}{\partial t} = -AM_{\alpha_{(s)}} k_r (C_{R_{(aq)}} - C_{P_{(aq)}} / K_{eq}) \tag{5}$$

where $A$ is the reactive surface area and $M$ is the molar volume.

## 3.2 Numerical method

### 3.2.1 LB fluid flow model

The LBM considers flow as a collective behavior of pseudo-particles residing on a mesoscopic level, and solves the Boltzmann equation using a small number of velocities adapted to a regular grid in space. Due to its underlying kinetic nature, the LBM is particularly useful in fluid flow applications involving interfacial dynamics and complex boundaries, e.g., multiphase or multi-component flows in porous media [*Chen et al.*, 2014b]. In the present study, incompressible LB model proposed by Guo et al. [2000] is employed for fluid flow which can eliminate the compressible effects of conventional LB model. The evolution of LB equation for fluid flow in the incompressible LB model is as follows

$$f_i(\mathbf{x}+\mathbf{c}_i\Delta t, t+\Delta t) - f_i(\mathbf{x},t) = -\frac{1}{\tau_\upsilon}(f_i(\mathbf{x},t) - f_i^{eq}(\mathbf{x},t)) \tag{6}$$

where $f_i(x,t)$ is the density distribution function, $f^{eq}$ is the $i$th equilibrium distribution function. $\tau_v$ is the collision time related to the kinematic viscosity. For the D2Q9 (two-dimensional nine-velocity) model used in this study as shown in Fig. 1, the discrete lattice velocity $\mathbf{c}_i$ is given by

$$\mathbf{c}_i = \begin{cases} 0 & i=0 \\ (\cos[\frac{(i-1)\pi}{2}], \sin[\frac{(i-1)\pi}{2}]) & i=1,2,3,4 \\ \sqrt{2}(\cos[\frac{(i-5)\pi}{2}+\frac{\pi}{4}], \sin[\frac{(i-5)\pi}{2}+\frac{\pi}{4}]) & i=5,6,7,8 \end{cases} \quad (7)$$

The main difference between incompressible LB model and conventional LB model lies in the equilibrium distribution function. For the incompressible LB model, the equilibrium distribution is defined as

$$f_i^{eq} = \begin{cases} -4\sigma \frac{p}{c^2} + s_i(\mathbf{u}) & (i=0) \\ \lambda \frac{p}{c^2} + s_i(\mathbf{u}) & (i=1-4) \\ \gamma \frac{p}{c^2} + s_i(\mathbf{u}) & (i=5-8) \end{cases} \quad (8)$$

where $\sigma$, $\lambda$ and $\gamma$ are parameters and satisfying the following equations:

$$\begin{cases} \lambda + \gamma = \sigma \\ \lambda + 2\gamma = \frac{1}{2} \end{cases} \quad (9)$$

and $s_i(\mathbf{u})$ is expressed as:

$$s_i(\mathbf{u}) = \omega_i \left[ 3\frac{\mathbf{e}_i \cdot \mathbf{u}}{c} + 4.5\frac{(\mathbf{e}_i \cdot \mathbf{u})^2}{c^2} - 1.5\frac{\mathbf{u} \cdot \mathbf{u}}{c^2} \right] \qquad (10)$$

with the weights $\omega_i=4/9$, $i=0$; $\omega_i=1/9$, $i=1,2,3,4$; $\omega_i=1/36$, $i=5,6,7,8$. Fluid density $\rho$ and velocity $\mathbf{u}$ can be obtained by

$$\mathbf{u} = \sum_{i=1}^{8} \mathbf{e}_i f_i, \quad \frac{p}{\rho} = \frac{c^2}{4\sigma}[\sum_{i=1}^{8} f_i + s_0(u)] \qquad (11)$$

The kinematic viscosity in lattice units is related to the collision time by

$$\upsilon = c_s^2(\tau_\upsilon - 0.5)\Delta t \qquad (12)$$

where $c_s$ is the speed of sound in the LB model. Eqs. (6) and (8) can recover the NS equations 3(a-b) using the Chapman–Enskog expansion [*Chen et al.*, 2013a].

*3.2.2 LB solute transport model*

For solute transport described by Eq. 3(c-d), the LB equation is written as

$$g_{i,k}(\mathbf{x}+\mathbf{c}_i\Delta t, t+\Delta t) - g_{i,k}(\mathbf{x},t) = -\frac{1}{\tau_{g,k}}(g_{i,k}(\mathbf{x},t) - g_{i,k}^{eq}(\mathbf{x},t)) \qquad (13)$$

where $g_{i,k}$ is the distribution function with velocity $c_i$ at the lattice site **x** and time $t$ for $k$th component, and $k = \alpha$ and $\beta$ for the main mineral $\alpha$ and the discrete mineral $\beta$, respectively. Note that the convection-diffusion equation (Eq. 3(c-d)) is linear in velocity **u**. This means that the equilibrium distributions for scalar transport need only to be linear in **u**; and thus lattices with fewer vectors is sufficient for scalar transport. Thus, a reduced D2Q5 lattice model is used in the present study (The lattice velocities $\mathbf{c}_i$=5,6,7,8 are not considered) and is combined with an equilibrium distribution that is linear in **u** [*Sullivan et al.*, 2005]

$$g_{i,k}^{eq} = \varphi(J_{i,k} + \frac{1}{2}\mathbf{c}_i \cdot \mathbf{u}) \qquad (14)$$

$J$ is given by [*Sullivan et al.*, 2005]

$$J_{i,k} = \begin{cases} J_{0,k}, & i = 0 \\ (1-J_{0,k})/4, & i = 1,2,3,4 \end{cases} \qquad (15)$$

where the rest fraction $J_0$ can be selected from 0 to 1. In literature, different forms of equilibrium distribution are adopted. The equilibrium distribution function given by Eq. (14) is a general formula, which becomes the one used in Huber et al. [2008] if $J_0$=1/3 and Kang et al. [2007] if $J_0$=0. The accuracy and efficiency of the reduced D2Q5 model has been confirmed in literature[*Chen et al.*, 2012b; *Chen et al.*, 2012c; *Chen et al.*, 2013a; *Chen et al.*, 2014a; *Chen et al.*, 2012e; *Huber et al.*, 2008; *Kang et al.*, 2007; *Luan et al.*, 2012]. Eq. (14) covers a wide range of diffusivity by adjusting $J_0$ in Eq. (15), which is a prominent advantage of such an equilibrium distribution [*Chen et al.*, 2012b; *Chen et al.*, 2013b; *Sullivan et al.*, 2005]. The concentration is obtained by

$$C = \sum g_{i,k} \tag{16}$$

The diffusivity is related to the collision time by

$$D_k = \frac{1}{2}(1 - J_{0,k})(\tau_{g,k} - 0.5) \tag{17}$$

Eqs. (13)-(14) can be proved to recover Eq. 3(c-d) using the Chapman–Enskog expansion [*Chen et al.*, 2013a].

*3.2.3 Boundary conditions in LB*

Unlike conventional numerical methods where the fundamental variables are macro physical quantities such as density, velocity, temperature and concentration, in the LB framework, the fundamental variables are the distribution functions. Therefore, boundary conditions based on the distribution functions are required when performing LB simulations. The LB boundary conditions used in the present study are briefly introduced below. For details of the boundary conditions, one can refer to the corresponding literature.

In the present study, for fluid flow two kinds of boundary conditions are adopted: a no-slip boundary condition at the fluid-solid interface, and a pressure boundary condition at the inlet and outlet of the computational domain. In the LB framework, for a no slip boundary condition, the bounce-back scheme is employed; and for a pressure boundary condition, the non-equilibrium extrapolation method proposed in Ref. [*Guo et al.*, 2002] is adopted.

For solute transport, four types of boundary conditions are used: the concentration boundary condition for the inlet, the outflow boundary condition for the outlet, the no-flux boundary condition for nonreactive solid surface and the reaction boundary condition described by Eq. (4) for reactive solid surface. In the LB framework, for the concentration boundary condition, when the D2Q5 lattice model is used only one distribution function is not known and is obtained by subtracting the concentration from the other four known distribution functions; for the outflow boundary condition, the unknown distribution function is simply set as the distribution function at the same direction of the neighboring fluid node; for the no flux boundary condition, bounce-back scheme is adopted; finally, for the reactive boundary condition described by Eq. (4), the boundary condition proposed in our previous work is employed [*Kang et al.*, 2007], which can guarantee the mass conservation very well.

*3.2.4 Update of solid phase*

Eq. (5) is updated at each time step based on the time forward difference

$$V_{\alpha_{(s)}}(t+\Delta t) = V_{\alpha_{(s)}}(t) - AM_{\alpha_{(s)}} k_r (C_{R_{(aq)}} - C_{P_{(aq)}} / K_{eq})\Delta t \qquad (18)$$

The method proposed by Kang et al. [2006] is used to update the structures of the solid phase. Once the mass of a solid node, $m_{\alpha_{(s)}}$, reaches zero, this solid node dissolves, becoming a new fluid node. Initialization of information related to fluid flow and mass transport for this new fluid node are implemented using the schemes proposed in Chen et al. [2013b]. Since this new fluid node is directly adjacent to solid phase, its initial velocity is set as zero. The pressure at this node is set as the averaged pressure of the nearest-neighbor fluid nodes. It is worth mentioning that in our previous studies density was initialized [*Chen et al.*, 2013b]. This is because in the incompressible LB model used in the present study (see Section 3.2.1), pressure rather than density is considered and the fluid density is a constant [*Guo et al.*, 2000]. Concentrations at this new fluid node also take the averaged concentrations of the nearest-neighbor fluid nodes.

*3.2.5 Numerical procedure*

After initialization, each time stepping involves the following sub-steps: (1) updating the flow field using the fluid flow LB model, then calculating the permeability and the hydraulic aperture; (2) solving the solute transport in the fluid with the dissolution reaction at the fluid-mineral α interface using the LB mass transport model, and (3) evolving the mass of solid nodes, updating the geometry of the solid phases and calculating the porosity, surface area, mechanical aperture and chemical aperture. Repeating (1)-(3) until the dissolution reaction completes. In our simulation, we assume the time scale of fluid flow is much shorter than that of the change of solid structures. In other words, the changes in porosity must occur over a time scale longer than the response time of the flow field to changing boundary conditions. This condition is met in our numerical simulations.

The numerical model has been validated by several typical fluid flow, mass transport and reactive transport processes in our previous studies [*Chen et al.*, 2012b; *Chen et al.*, 2013a; *Chen et al.*, 2013b; *Chen et al.*, 2012e; *Kang et al.*, 2007]. For brevity, these validations are not repeated here and one can refer to these studies for more detail.

## 4. Dissolution in fractured porous media

The fractured porous medium used in this study is shown in Fig.2. It consists of a fracture sandwiched by two rectangular rocks. The fracture aperture is 400μm and the height of each rock is 300μm. The volume fraction of mineral β in the rock is 0.2. The entire domain has a size of 4000*1000μm and is discretized using 400*100 lattices with lattice resolution of 10 μm. Five additional lattices are added at the left and right boundary, respectively, to avoid the effects of the solid phase on the fluid flow at the boundaries.

Initially, the fracture is filled with a fluid without $R_{(aq)}$ and $P_{(aq)}$. A constant pressure gradient is applied at both left and right boundaries, driving the fluid flow from inlet to outlet. After flow becomes steady, $R_{(aq)}$ with a constant concentration $C_{R,in}$ is injected from the left inlet, but the concentration of $P_{(aq)}$ is fixed at zero at the left inlet. Then dissolution reaction takes place at the fluid-mineral α interface, consuming $R_{(aq)}$ and α as well as producing $P_{(aq)}$. Dissolution processes strongly depend on physicochemical conditions. The relative strength of convection, diffusion and reaction rate can be well represented by two important dimensionless numbers, namely the Peclet number and Damkohler number. The Peclet number, *Pe*, measures the relative magnitude of convection and diffusion

$$Pe = \frac{\bar{u} l_c}{D} \tag{19}$$

where $\bar{u}$ is the averaged velocity in the system and $l_c$ is the characteristic length chosen as the initial aperture of the fracture. The Damkohler number, *Da*, stands for the ratio between reaction and diffusion

$$Da = \frac{k l_c}{D} \tag{20}$$

Here the definition of *Da* follows that in a series of recent publications [*Lichtner and Kang*, 2007; *Alexandre M. Tartakovsky et al.*, 2007; *Tartakovsky et al.*, 2008]. Note that for reactive transport processes with dissolution, the *Pe* and *Da* are changing because both the flow rate and fracture aperture vary with time due to the dissolution. In the present study, we discuss the simulation results using the initial *Pe* and *Da*. In the flow regimes considered in the present study, the inertial force can be neglected and the *Re* number is very small and its value is not important.

The parameter values in the present study are listed in Table 1, some of which depend on the simulation cases. For example, although the diffusivity in physical unit is fixed, it varies in the LB framework to obtain a wide range of $Pe$ values. As can be seen in Eq. (19), to achieve a high $Pe$, one can either use a large $\bar{u}$ or a small $D$. However, in the LB simulation of fluid flow, the low Mach number condition ($Ma<<1$) must be met to ensure the simulation accuracy. Since the speed of sound in the D2Q9 lattice used in the current study is 0.577 (in lattice units), the velocity in the LB simulations cannot be very high. Therefore, to achieve a high $Pe$ while satisfying the low Mach number condition, the diffusion coefficient $D$ in lattice units needs to be reduced. According to Eq. (17), this can be done by either decreasing the collision time $\tau_g$ or increasing the rest fraction $J_0$. It is worth mentioning that different $\tau_g$ leads to different physical time for each time step in the LB simulation, when the physical diffusivity is fixed. The lower the $\tau_g$, the shorter the physical time for each time step, and the larger the required simulation steps. So care needs to be taken when converting variables between physical units and lattice units. In the following paragraphs, the values of $\Delta p$ and $\tau_g$ ($J_0$ is fixed as 0.2 in the present study) are given for each simulation case. Similarly, the effective forward reaction rate constant $k_r$ is also varied to obtain a wide range of $Da$ values.

In the LBM, the variables are in the lattice units instead of physical units. To implement the LB model it is necessary to obtain a consistent set of parameters that correspond to the physical parameters for the problem at hand [*Lichtner and Kang*, 2007]. This is achieved by equating various dimensionless groups. For diffusivity

$$\frac{D_P}{\Delta x_P^2 / \Delta t_P} = \frac{D_L}{\Delta x_L^2 / \Delta t_L} \tag{21}$$

with subscript "P" and "L" representing "physical units" and "lattice units", respectively. $\Delta x_P$ and $\Delta t_P$ equal unit in the LBM. After the computational domain is defined and discretized, the physical length of a lattice $\Delta x_P$ can be determined. Therefore, with the diffusivity in physical units given in Table 1 and the diffusivity in lattice units determined according to Eq. (17) with $J_0$ and $\tau_g$, the time scale $\Delta t_P$ can be calculated according to Eq. (21). Other variables in physical units, such as reaction rate coefficient $k_r$, can be calculated from the their corresponding one in lattice units by matching the dimensionless number

$$\frac{k_{r,P} l_P}{D_P} = \frac{k_{r,L} l_L}{D_L} \tag{22}$$

For transfer concentration in different units, $C_P/C_{0P} = C_L/C_{0L}$ is used, with $C_{0P} = 1000$ mol m$^{-3}$ and $C_{0L} = 1$.

**4.1 Effects of undissolved mineral**

*4.1.1 Dissolution morphology*

Several simulations are conducted for *Pe* within (0.011, 5.5) and *Da* within (0.01, 5). Three cases are selected from the simulation results to represent the typical reactive transport processes and dissolution patterns: (1) *Pe*=0.011, *Da*=0.01; (2) *Pe*=0.011, *Da*=5; and (3) *Pe*=5.5, *Da*=5, where (1) $\Delta p$=0.0001, $\tau_g$=1.0; (2) $\Delta p$=0.0001, $\tau_g$=1.0; and (3) $\Delta p$=0.01, $\tau_g$=0.6, respectively. They represent the reaction-controlled process with low flow rate, diffusion-controlled process with low flow rate and diffusion-controlled process with high flow rate, respectively. Note that for the reaction-controlled case, whether the flow rate is high or low does not make much difference [*Békri et al.*, 1995]. The temporal evolutions of the concentration distributions of $R_{(aq)}$ as well as the geometries of the rocks for the three cases are presented in Figs. 3-5. The colored contours in the void space denote solute concentration. For each case, three images are shown representing the initial stage, the middle stage and near-end stage. The concentration is normalized by the inlet concentration of $R_{(aq)}$. Generally, the dissolution process is determined by the concentration of $R_{(aq)}$ at the solid-fluid interface. From the set of images of in Figs. 3-5, it can be seen that the *Pe* and *Da* have significant effects on the mass transport and geometrical morphology and structure evolution of the rock. The dissolution of mineral α roughly shows the characteristic of relatively uniform dissolution (*Pe*=0.011, *Da*=0.01), face dissolution (*Pe*=0.011, *Da*=5) and fracture dissolution (*Pe*=5.5, *Da*=5), corresponding to the familiar phenomena compared with previous simulation of dissolution of rocks with mono-mineral [*Békri et al.*, 1995; *Huber et al.*, 2013; *Kang et al.*, 2003]. For (*Pe*=0.011, *Da*=0.01), the process is reaction-controlled, indicating the diffusion is much faster than the reaction, leading to very uniform distributions of $R_{(aq)}$ in the entire void space and thus uniform dissolution of mineral α, as shown in Fig. 3 (a). The *Pe* number in the second case is the same to the first case, but *Da* number is increased to 5, indicating that diffusion is slow compared to reaction, leading to diffusion-controlled process. Due to the slow diffusion as well as the low flow rate, most of $R_{(aq)}$ is consumed at the reaction front and the downstream region of the fracture is starved of $R_{(aq)}$. Correspondingly, dissolution mainly takes place at the α-fluid interface facing the inlet boundary (Fig. 4(a)). The dissolution front gradually advances towards the downstream (Fig. 4(b-c)), behind which the fracture-rock surface is intact. Obviously, solute transport in the above two cases is mainly by diffusion as the *Pe* is very low. In the third case the *Pe* is increased to 5 while keeping *Da*=5, leading to diffusion-controlled process with fast flow rate. The dominated mechanism of mass transport now is convection, meaning concentration of $R_{(aq)}$ is higher in the main flow path, namely in the fracture. Thus, the dissolution advances along the fracture-rock interface, as shown in Fig. 5.

While in general the dissolution processes in Figs. 3-5 are similar to that in rock with mono-mineral, the existence of the undissolved β significantly affects the local dissolution behaviors. For the (*Pe*=0.011, *Da*=0.01) case, although we use to the term "uniform dissolution" to describe the dissolution characteristic, the dissolution can only be considered as uniform at the global scale, as it is extremely un-uniform at the local scale. This is in agreement with the experimental observations of dissolution in rocks with undissolved minerals: although the dissolution is uniform at sample scale, it shows microscopically strong heterogeneity [*Noiriel et al.*, 2013]. The undissolved mineral β leads to the local heterogeneity of the dissolution (Figs. 3-5). A porous layer formed by β remains between the main fracture and dissolution front as shown in Figs.3-5. The formation of such a porous layer due to undissolved minerals was also observed in experiments [*Noiriel et al.*, 2007]. This porous layer acts as a barrier for mass transport. Diffusion through it slows down or even is not allowed where mineral α is heavily surrounded by β (one of them is shown in the circle of Fig. 3 (b)). These dissolution-forbidden regimes present as bulges on the rock-fluid interface during the dissolution process as shown in Fig. 3-5, clearly demonstrating the effects of the undissolved mineral β on the entire dissolution patterns. The rock surface under dissolution is rather rough compared with that in rocks with mono-mineral where the dissolving rock surface is very smooth [*Kang et al.*, 2003]. Furthermore, in the case with $Pe$=5.5 and $Da$=5, the gradient of concentration of $R_{(aq)}$ in the porous layer is much higher than that in the fracture. This is because the dominant mass transport mechanism inside the porous layer is diffusion, while that in fracture is convection. As the characteristic time of diffusion is much longer than that of convection, concentration gradient inside the porous media is more prominent, further highlighting the barrier effects of the porous layer on mass transport.

*4.1.2 Porosity and permeability*

Now attention is turned to effects of undissolved mineral on the porosity $\varepsilon$ and permeability $k$. The initial porosity of the domain $\varepsilon_0$ is 0.40 and the permeability $k_0$ is 50.0 in lattice units. With the resolution of one lattice as $1\times10^{-5}$m, the permeability in physical units is $5.0\times10^{-9}$ m$^2$. The porosity and permeability in Fig. 6(a) are normalized with their initial values. In Fig. 6(a), the case of $Pe$=0.11 and $Da$=5, corresponds to $\Delta p$=0.001 and $\tau_g$=1.0; the case of $Pe$=0.11 and $Da$=0.01, corresponds to $\Delta p$=0.001 and $\tau_g$=1.0; and the case of $Pe$=1.1 and $Da$=5 corresponds to $\Delta p$=0.01 and $\tau_g$=1.0. Note that the rock is located with (50μm, 4050μm), while the porosity and permeability are calculated within (40μm, 4060μm), with one additional lattice (10 μm) at both ends to obtain the exact pressure drop across the rock.

Several typical characteristics can be observed in Fig. 6(a). First, all of the simulation cases stop at $\varepsilon/\varepsilon_0$=2.2, indicating that amount of dissolution of mineral α for all the cases is the same. As the volume fraction of mineral β in the rock is 0.2, it can be further concluded that mineral α is completely dissolved for all the cases. $k/k_0$ for all the cases also ceases at the same value (about 1.23). This is expected as there is no difference on the ultimate porous structures of the rock for different cases.

Second, for the same *Da*, increasing *Pe* can raise the permeability more efficiently. This is because higher *Pe* (fracture dissolution) can expand the entire fracture while lower *Pe* (face dissolution) only leads to local dissolution. Besides, the (*Pe*=0.11, *Da*=0.01) case is much close to the curve $y = x^3$. Here we want to make a few comments on the curve $y=x^3$. It is known that fluid flow between two parallel static walls is called Poiseuille flow and an analytical solution exists for the permeability $k_h$ [*Kang et al.*, 2002a], namely $k_h = h^2/12$, where *h* is the distance between the two walls or is the fracture aperture in the present study. Therefore, the permeability of the domain simulated is $k_H = (h^2/12) \cdot (h/H)$. $k_H$ normalized by its initial value $k_H^0$ can be written as

$$\frac{k_H}{k_H^0} = \left(\frac{h/H}{h_0/H}\right)^3 = \left(\frac{\varepsilon}{\varepsilon_0}\right)^3 \quad (23)$$

where $\varepsilon = h/H$ is the porosity of the fractured medium. Eq. (23) indicates that the normalized permeability and porosity of the fractured medium obey a power law with an exponent of 3. Eq.(23) is only valid when aperture growth due to dissolution is uniform along the entire fracture. The (*Pe*=0.11, *Da*=0.01) case is quite close to this scenario, but the undissolved β in the system leads to the discrepancy.

Third, the effect of the undissolved mineral on the porosity-permeability relationship is evident. For comparison, we also simulated the dissolution processes for the mono-mineral rock, and the porosity-permeability relationship under different *Pe* and *Da* is plotted in Fig. 6(b). All the initial and boundary conditions are the same as that in Fig. 6(a) except that the rock consists of mineral α only. In Fig. 6(b), for (1) *Pe*=0.011, *Da*=5; (2) *Pe*=0.11, *Da*=5; (3) *Pe*=1.1, *Da*=5; ;(4) *Pe*=5.5, *Da*=5 and (5) *Pe*=0.11, *Da*=0.01, the corresponding pressure different and collision time are (1) $\Delta p$=0.0001, $\tau_g$=1.0; (2) $\Delta p$=0.001, $\tau_g$=1.0; (3) $\Delta p$=0.001, $\tau_g$=0.55; (4) $\Delta p$=0.001, $\tau_g$=0.51; and(5) $\Delta p$=0.001, $\tau_g$=1.0;

As can be seen in Fig. 6(b), for mono-rock mineral rock dissolution, the curves for all the cases are concave. This indicates that dissolution is increasingly efficient to enhance the permeability as the reactive transport process proceeds, which is because the transport of $R_{(aq)}$ becomes more efficient as the permeability increases due to dissolution. Note that the final $k/k_0$ for all the case should equal $(1/\varepsilon_0)^3$ as the final porosity $\varepsilon = 1$. This is exactly the case in Fig. 6(b), where all the curves end at $\varepsilon/\varepsilon_0 = 2.52$ and $k/k_0$=16.0, leading to good coincidence of the simulation curves with $y = x^3$ at $\varepsilon/\varepsilon_0 = 2.52$. The slight discrepancy is attributed to the additional lattices included on each boundary of the rock to calculate the porosity and permeability, as mentioned previously.

It can be seen that the curves in Fig. 6(a) are very different from that in Fig. 6(b). In Fig. 6(a), the permeability reaches a steady state when the porosity approaches a certain value ($\varepsilon/\varepsilon_0$)$_c$, indicating that dissolution after the stage with a porosity higher than ($\varepsilon/\varepsilon_0$)$_c$ does not contribute to the increase of the permeability. For the same *Da*, ($\varepsilon/\varepsilon_0$)$_c$ decreases as *Pe* rises. The reason is because the fluid mainly flows in the fracture rather than in the porous layers where flow resistance is high. Hence, the dissolution of α in the deep porous layer of mineral β has little contribution to the permeability. This is also found in the experiments of Noiriel et al. [2007]. Besides the experimental study in Noiriel et al. [2007] even found that the permeability would drop when the porosity increases as the fracture is blocked by the undissolved clay particles flowing with the fluid. Sometimes the fractures would collapses due to the mechanical effects [*Polak et al.*, 2004]. In the present study, the remaining β particles are fixed at their initial positions, and do not move with the surrounding fluid. This assumption is reasonable if we consider the computational domain as a two-dimensional slice of a three-dimensional structure in which the mineral β is connected. Under such scenario, the mineral β remains stationary due to the connection. In fact, it is a common practice to perform pore-scale simulations based on two-dimensional slices where the solid particles are disconnected or even scattered [*Kang et al.*, 2002b; *Kang et al.*, 2014; *Li et al.*, 2008; *Alexandre M. Tartakovsky et al.*, 2007; *Tartakovsky et al.*, 2008]. Such 2D simulations can greatly reduce the computational resources required for the 3D simulations and are still capable of revealing the underlying mechanisms of the physicochemical processes.

In Fig. 6(a), for $\varepsilon/\varepsilon_0$ equal to 2.0, $k/k_0$ only equals 1.23. However, the permeability increases substantially with $k/k_0$=16.0 for $\varepsilon/\varepsilon_0 = 2.52$ in Fig. 6(b). The final permeability in Fig. 6(a) is much lower than that in Fig. 6(b) for the same amount of dissolution. It again demonstrates the significantly negative effects of the undissolved mineral on the increase of permeability. Hence, conventional widely used porosity-permeability empirical relationships, such as the KC equation [*Carman*, 1937], will significantly overestimate the permeability when undissolved mineral exists in a system, and one must be very careful when applying such an empirical relationship to predict reactive transport processes in a system with undissolved minerals. Finally, it can be seen in Fig. 6(a) and 6(b) that the permeability is not generally a simple function of the porosity, which strongly depends on the flow velocity, mass transport and reaction rate. As a result, further efforts are required to investigate the relationships between the hydrological properties and the structural parameters of the porous media under different reactive transport conditions [*Luquot and Gouze*, 2009].

Here, an analytic solution for the permeability of the fractured system consisting of an open region sandwiched by two homogeneous porous media (shown in Fig. A1) is presented. In fact, fluid flow along the interface of a porous medium and an open flow region is a common phenomenon often seen in engineering applications, such as fluid flow in porous medium enhanced heat exchanger, fuel cells and micro reactors [*Beavers and Joseph*, 1967 ; *Cao et al.*, 2013; *Chen et al.*, 2012d], as well as fluid flow in fractures as studied in the present study. For such a fractured system, the analytical solution for the permeability is given by

$$k = \frac{1}{h+2h_\mathrm{p}}(\frac{h^3}{12}+\frac{1}{2}\frac{\sqrt{k_\mathrm{p}}}{\alpha}h^2+k_\mathrm{p}h+2k_\mathrm{p}h_\mathrm{p}) \qquad (24)$$

For the detailed derivation, one can refer to the Appendix. If the porous medium is impermeable, namely, its permeability $k_\mathrm{p}$ is zero, Eq. (22) is reduced to

$$k_0 = \frac{\frac{h^3}{12}}{h+2h_p} \tag{25}$$

which is exactly the initial permeability of the fractured medium as discussed in Eq. (23). The ratio between $k$ and $k_0$ is

$$\frac{k}{k_0} = 1 + 6\frac{\sqrt{k_p}}{\alpha}h^{-1} + 12k_p h^{-2} + 24k_p h_p h^{-3} \tag{26}$$

To calculate $k/k_0$, the permeability of the porous medium, $k_p$ is needed. It can be obtained by simulating fluid flow in the porous medium at the pore-scale using the LBM introduced in the present study. Here, the KC equation is employed to calculate $k_p$ based on the porosity $\varepsilon_p$

$$k_p = c\frac{\varepsilon_p^3}{(1-\varepsilon_p)^2} \tag{27}$$

where $c$ is a geometric factor calculated by $c=d^2/180$. Based on the structural characteristics of mineral β in Fig. 4, the particle diameter $d$ is set as 5 lattices, namely 50μm. When the dissolution is completed, $\varepsilon_p$ =0.8, as volume fraction of the undissolved mineral β is 0.2. Consequently, $k_p$ can be calculated and is equal to $2.56 \times 10^{-12}$ m$^2$. Another variable required to calculate $k/k_0$ is the dimensionless quantity α, which depends on the structures of the porous medium [*Beavers and Joseph*, 1967]. α is assumed to be 1.0 [*Beavers and Joseph*, 1967] in the present study. Finally, with $k_p$ and α determined, substituting the height of the fracture ($h$=400μm) and the porous medium ($h_p$ =300μm) in Eq. (26), $k/k_0$ is calculated as 1.218, in good agreement with the value 1.23 predicted by our pore-scale simulation.

*4.1.3 The aperture*

In experiments, monitoring the aperture change is an assistive method for understanding the reactive transport processes taking place inside the sample. There are different fracture apertures determined based on different aspects, including chemical aperture, hydraulic aperture and mechanical aperture [*Noiriel et al.*, 2007; *Noiriel et al.*, 2013]. In experiments, the chemical aperture is calculated on the basis of mineral mass balance. The outlet concentrations of interested ions $C_{\text{ion}}$ are monitored and are used to inversely calculate the dissolving volume of the sample [*Luquot and Gouze*, 2009; *Noiriel et al.*, 2007; *Noiriel et al.*, 2013]. The chemical aperture $a_c$ is then determined by

$$\frac{da_c}{dt} = -\frac{1}{A}\frac{dV_{\text{rock}}}{dt} = -Qv_{\text{ion}}C_{\text{ion}} \qquad (28)$$

where $V_{\text{rock}}$ is the volume of the rock and $A$ is the equivalent area of the fracture wall; $v_{\text{ion}}$ is the molar volume of the mineral under dissolution; and $Q$ is the volumetric flow rate. However, in the present pore-scale study, the volume variations can be tracked in real-time, a distinct advantage of pore-scale simulations which can track the temporal evolutions of pore-scale structures. Hence, the inverse calculation of the volume variation from the outlet concentrations is not required and the chemical aperture $a_c$ at time $t$ can be directly calculated by

$$a_c = a_0 - \frac{1}{A}(V_{rock,t} - V_{rock,0}) = a_0 - \frac{V}{A}(\varepsilon_0 - \varepsilon_t) = a_0 - H(\varepsilon_0 - \varepsilon_t) \qquad (29)$$

where $V$ is the total volume of the fracture including the solid phase and void space, and $H$ is the height of the computational domain. As can be seen from Eq. (29), by recording the porosity during the simulation, the chemical aperture can be directly calculated.

Another important aperture is the hydraulic aperture $a_h$, which is measured throughout the flow-through experiment by recording the pressure drop $\Delta p$ along the sample [*Luquot and Gouze*, 2009]. The hydraulic aperture $a_h$ is calculated by the following expression using the cubic law [*Zimmerman and Bodvarsson*, 1996]

$$a_h = \sqrt[3]{\frac{12\mu L Q}{\Delta p l}} = \sqrt{\frac{12\mu L \bar{u}}{\Delta p}} \qquad (30)$$

where $L$ is the length of the aperture. $l$ is the width of the fracture in the $z$ direction (the third dimension neglected in the present 2D simulation). On the other hand, according to the Darcy's law, the permeability is determined by

$$k_{\text{h}} = \frac{\mu L \bar{u}}{\Delta p} \qquad (31)$$

Therefore, the hydraulic aperture $a_{\text{h}}$ is related to the permeability by [*Noiriel et al.*, 2007]

$$a_{\text{h}} = \sqrt{12 k_{\text{h}}} = \sqrt[3]{12 k_{\text{H}} H} \qquad (32)$$

During our simulation, $k_{\text{H}}$ is recorded and its variation is discussed in Section 4.1.2. $a_{\text{h}}$ is calculated from $k_{\text{H}}$ according to Eq. (32).

Last but not least, the mechanical aperture $a_{\text{m}}$ is another important aperture which is obtained by directly measuring the aperture based on the fracture morphology [*Noiriel et al.*, 2013]. In the present study, the following algorithm is employed to determine the mechanical aperture: (1) Starting from the central cell at $x=50\mu m$ and $y = H/2$ (a void space cell as shown in Fig. 2), moving perpendicularly towards the top boundary until encountering the first solid cell, the distance between the central cell and this solid cell is recorded as $h_{1,x}$; similarly, starting from the central cell, perpendicularly towards the bottom boundary until encountering the first solid cell, the distance between the central cell and this solid cell is recorded as $h_{2,x}$. (2) The local mechanical aperture $a_{m,x}$ is determined as the sum of $h_{1,x}$ and $h_{2,x}$. (3) Repeating (1) and (2) along $x$ direction until all the local mechanical aperture $a_{m,x}$ is obtained. The mechanical aperture $a_m$ is chosen as the averaged value of the $a_{m,x}$ with $x$ in the range (50μm, 4050μm). By this algorithm only the widest void space at any $x$ position is chosen when the fracture walls locally overlap [*Noiriel et al.*, 2013]. Fig. 7(a) shows the distributions of $a_{m,x}$ when simulation is converged, which is fluctuant due to the complex local dissolution processes as shown in Fig. 5.

Fig. 7(b) plotted the changes of the hydraulic, mechanical and chemical apertures for $Pe$=1.1 and $Da$=5. Discrepancy between the different apertures is considerable. The chemical aperture rises remarkably from 400μm to 875μm, in line with the change from an initial porosity of 0.4 to a final porosity of 0.875 according to Eq. (29). Although there has been significant dissolution, the hydraulic aperture changes slightly from 400μm to 428.5μm, indicating slight enhancement of fluid flow. This is expected according to Eq. (32) as the domain permeability $k_H$ only increases 1.23 times as discussed previously due to the formation of porous layers formed by the undissolved mineral. The mechanical aperture lies between the chemical aperture and the hydraulic aperture with an increment of about 137μm. This is because only part of the dissolution contributes to the increase of the local aperture. Obviously, different apertures reflect different aspects of the dissolution processes. The chemical aperture indicates the dissolution amount, the hydraulic aperture stands for the contribution of the dissolution to the fluid flow, and the mechanical dissolution describes the overall morphology of the fractures. The large discrepancy shown in Fig. 7 clearly demonstrates the heterogeneous characteristics of the dissolution. Therefore, these three apertures usually supplement each other to better understand the reactive transport processes during dissolution in experiments [*Noiriel et al.*, 2007].

### 4.2 Effects of particle structure of mineral β

In this section, effects of particle structures of the undissolved mineral on the dissolution as well as hydraulic properties are investigated. Three structures of the mineral β particle, namely, random structure (Fig. 2(a)), *x*-oriented structure (Fig. 2(b)) and *y*-oriented structure (Fig. 2(c)) are simulated. For all the three cases, $Pe$=1.1, $Da$=5 ($\Delta p$=0.01 and $\tau_g$=1.0) and the volume fraction of mineral β in the rock is the same. Fig. 8 displays the relationship between porosity and permeability. It can be seen that *x*-oriented particle generates the largest ultimate permeability, followed by random-structures and then *y*-oriented structures. This is because the *y*-oriented particles are perpendicular to the main flow reaction, namely in the *x* direction, causing the largest flow resistance in the porous layer. Therefore, the permeability only increases 1.17 times, although 80% of the volume of the rock is dissolved. This indicates that the morphology of the undissolved mineral plays an important role on the hydraulic properties of the rock under dissolution. While in the present study the morphology of the mineral β is prescribed initially and remains unchanged during the simulation, reorganization of the structures of undissolved mineral may happen due to the surrounding fluid flow and loss of the support of the dissolved mineral [*Noiriel et al.*, 2007]. Such reorganization certainly changes the particle structures of the undissolved mineral, leading to different morphologies such as those studied in this section. Obviously, hydraulic properties of the rock will also be dynamically affected by such reorganization, based on the results shown in Fig. 8.

### 4.3 Effects of dissolution rate of the mineral β

In this section, the mineral β is allowed to dissolve, but at a lower rate than mineral α. For all the simulation cases, $Pe=1.1$, $Da=5$ ($\Delta p=0.001$ and $\tau_g=0.55$) and the volume fraction of mineral β in the rock is the same as that Section 4.1. The simulation results for dissolution rate $k_2 = 1.0$ $k_1$, 0.1 $k_1$, 0.01 $k_1$ and 0 are shown in Fig. 9. Note that the cases for $k_2 = 0$ and $k_2 = 1.0$ are also shown in Fig. 6(a) and Fig. 6(b), respectively. Obviously, for $k_2$ greater than zero, the entire rock can be dissolved completely, and the final permeability and porosity are the same. However, before the final stage, for the same dissolution amount, the lower the dissolution rate of mineral β, the lower the permeability. This is because the lower the dissolution rate, the longer the mineral β exists, and thus the more significant the effects of the mineral β on the fluid flow. For the case with $k_2 = 0$, the reaction transport process ceases at a relatively lower porosity and a remarkably smaller permeability, which has been explained in Section 4.1.

## 5. Dissolution in rocks with complex porous structures

In this section, rocks with complex realistic porous structures are simulated, as shown in Fig. 10. The porous geometry consists of pieces of rocks of varying sizes surrounded by randomly oriented and interconnected micro fractures. QSDS described in Section 2 is employed to generate the distribution of mineral β with random distributions in the rocks with the structure controlling parameters as ($c_d=0.08$, $P_{1,2,3,4}=0.0004$, $P_{5,6,7,8}=0.0001$, $\varepsilon_\beta=0.3$). The domain size is 3000×3000μm and is discretized using 300×300 lattices. Twenty additional lattices are added at the left and right boundary, respectively, to avoid the effects of solid phase on the fluid flow at the boundaries. Typical width of the micro fractures is about 200μm. Boundary and initial conditions are the same as that in Section 4.1. Only one scenario is considered, namely $Pe$=6.01 and $Da$=5 ($\Delta p$=0.01 and $\tau_g$=0.55), to focus our study on the effects of the undissolved mineral on the formation of wormholes, an important phenomenon in both geosciences and engineering applications. The characteristic length used for calculating $Pe$ and $Da$ is the averaged radius of all the separated rocks. Here, the radius of a piece of rock is defined as the radius of a circle whose area is the same as the rock. The characteristic length calculated is 370μm. Mineral β is not allowed to dissolve in the simulations.

The temporal evolutions of the concentration distributions of $R_{(aq)}$ as well as the geometries of the rocks are displayed in Fig. 11. Reactant $R_{(aq)}$ from the left inlet enters the domain and flows into several fractures ($t$ = 2s and $t$=200s). Fingering-like contours with a sharp front in the main fracture clearly prove convection as the dominant mechanism of local mass transport. Dissolution of the mineral α at mineral α-fracture interfaces with local strong fluid flow can be clearly observed ($t$=200s), leaving porous layers formed by the undissolved mineral β, similar to the simulation results in Section 4. At $t$ = 200s, the solute penetrates the domain through an interconnected fracture with lowest tortuosity, marked as Arrow 1. It is well known that for dissolution in rocks with mono-mineral [*Békri et al.*, 1995; *Kang et al.*, 2003], positive feedback between the fluid flow and rock dissolution will take place. Dissolution enlarges the aperture, thus enhancing fluid flow and solute transport in the preferred fracture, which in turn reinforces its expansion. This phenomenon is called "wormholing" [*Daccord et al.*, 1993]. Fluid flow, solute transport as well as dissolution mainly occur in the preferred fracture. However, such positive feedback is somewhat suppressed when undissolved minerals exists. Due to the porous layer of the undissolved mineral, the increase of the fracture permeability is not significant. As a result, the fluid flow and solute transport in this fracture is not significantly enhanced, as discussed in Section 4. In the meantime, other fractures are opened as the reactive mineral is dissolved (marked as Arrow 2 and Arrow 3 at $t$ = 500s), competing with the initially preferred fracture for flow and solute transport. In other words, the dissipation of the concentration gradient built as well as the high flow resistance offered by the porous layer of the undissolved mineral breaks the positive feedback between the fracture expansion and enhanced flow and solute transport, and hence suppresses the formation of the wormhole in the initially preferred fracture. Such effects of the undissolved minerals is consistent with recent experiments [*Noiriel et al.*, 2007].

After several main fractures have been formed ($t$ = 500s), dissolution mainly takes place in the porous rocks and in the previously bypassed fractures ($t$=1000s~5600s). Dissolution in these regions is extremely slow as the local dominant transport mechanism is diffusion which is inefficient. Most of the reactants exit the domain from the right outlet, without reacting with the mineral. Such dissolution is inefficient to enhance the permeability of the domain. As shown in Fig. 12, after about $t$=500s, although the porosity keeps rising, the permeability quickly plateaus, and the reasons has been discussed in Section 4. Note that there is a sharp rise of the permeability at about $t$=500s. This is because the rocks in the circle marked at $t$ = 200s are completely dissolved at that time, leading to the opening of the path marked as Arrow 2 as shown at $t$ = 500s, causing the remarkable increase of the permeability. The porosity continues increasing but with a gradually reducing rate indicating the diminishing reactive surface area as well as a decrease in dissolution rate due to stronger mass transport resistance in the porous layer.

Changes in the rock volume due to dissolution also lead to the variation of surface area with time, as shown in Fig. 13. While porosity changes can be directly calculated from the dissolution amount, the surface area, which is related to the detailed morphology of the rocks, is more complex to determine. The morphology of rocks under dissolution is affected by several factors including flow rate, mass transport, chemical reactions as well as the structural characteristics of the rocks [*Noiriel et al.*, 2009]. As shown in Section 4, for the fractured porous media studied, which is a relatively simple structure, different *Pe* and *Da* lead to different dissolution patterns. Hence, the changes of the surface area are significantly affected by *Pe* and *Da*. In continuum models for reactive transport process, the reactive surface area is an important prerequisite for calculating the dissolution/precipitation reaction rate. However, it is not easy to measure during experiments. Several models based on geometrical constructions of the porous medium have been developed to relate surface area changes with the solid volume [*Emmanuel and Berkowitz*, 2005]. The sphere model is the simplest one which assumes that the porous medium consists of separated spherical particles. In this model, the surface area *A* is calculated by

$$A = A_{initial}(\frac{V}{V_{initial}})^{2/3} \tag{33}$$

Obviously, surface area decreases as the rocks dissolves. Alternatively, a porous medium can be assumed to consist of separated sphere pores. In this pore model, the surface area *A* is calculated by

$$A = A_{initial}(\frac{V}{V_{initial}})^{-2/3} \tag{34}$$

In Eqs. (33-34), $V$ is the volume of the solid phase. For sphere model, the surface area decreases as the solid volume reduces, while for the pore model, it increases as the solid volume decreases. Fig. 13 shows the time evolution of the total surface area of the rocks as well as the reactive surface area, namely the surface area of the mineral α. Calculated values using Eqs. (33-34) based on the volumes of the rock and mineral α are also plotted in the figure. The surface area calculated using the volume of the rock is the total surface area while that using volume of mineral α is the reactive surface area. Since the simulated total surface area and reactive surface area are obtained by directly counting the rock and mineral α boundary cells in the domain, respectively, they can be considered as the benchmark. Note that the area in Fig. 13 is normalized by the initial surface area of the mineral α, $A_0$. Initially, the ratio between the total surface area and $A_0$ is 1.23, indicating most of the rock surface is occupied by mineral α, consistent with the mineral distributions in Fig. 10. It can be seen that the simulated total surface area increases as the dissolution proceeds. The sphere model predicts the wrong change trend, while the pore model, although shows the correct change trend, underestimate the benchmark. Unlike the total surface area, the simulated reactive surface area, namely the surface area of mineral α, gradually decreases. Based on the observation of Fig. 10, it seems the mineral α can be approximately treated as isolated spherical particles. The reactive surface area calculated using the sphere model is in relatively good agreement with the simulated one. The reactive surface area calculated using the pore model, however, is very different from the numerically simulated one. This can be explained by a direction observation of the initial morphology of Fig. 10. Clearly the mineral α can be approximately treated as isolated spherical particles.

The considerable difference and the opposite change trend between the total surface area and the reactive surface area indicates that, the increase of the total surface area is not caused by the increase in the surface area of the reactive mineral, but by that of the unreactive mineral. When using surface area in the continuum model, the two surface areas must be clearly distinguished when unreactive minerals are present. Otherwise the simulation results will be even qualitatively wrong. While currently it is possible to measure the total surface area, the experimental determination of the reactive surface area is still challenging [*Noiriel et al.*, 2009]. The study in the present study demonstrates that the pore-scale simulations provide an efficient way to this end.

## 6. Conclusion

Effects of the undissolved mineral on the dissolution processes in fractured and complex porous media are studied in the present study. A pore-scale numerical model based on the LBM and CA model are employed to simulate the multiple physicochemical processes including fluid flow, mass transport, chemical reactions and modifications of solid phase involved in the dissolution processes. The distributions of minerals in the rocks are generated by a reconstruction method called QSGS.

Simulation results of dissolution process in a fractured porous medium reveal significant effects of the undissolved mineral on the reactive transport processes as well as the variations of porosity, permeability and apertures. As the dissolution proceeds, the undissolved mineral remains behind the dissolution front and forms porous layers between the fractures and the dissolution front. In the porous layers, mass transport rate is pretty low and flow resistance is very high. Hence, dissolution in the porous layers contributes little to the change in permeability. After the porosity reaches a certain value, the dissolution in the porous layers has no effects on the permeability which remains constant. Compared with the dissolution in rocks with mono-mineral, the enhancement of the permeability by dissolution is pretty low in the rocks with undissolved mineral. Due to the heterogeneity of the dissolution at local scale, the chemical, hydraulic and mechanical aperture show large discrepancies. Characterizing mass transfer rates through these fractured systems is critical for predicting hydrocarbon production rates in tight shales or evaluating if $CO_2$ can be sequestered under low permeability caprock. These results imply that the relationship between reaction, permeability and porosity are complex and must be characterized to properly simulate mass transfer in these systems.

Dissolution in rocks with complex porous structures is also investigated. It is found that the high flow resistance and the low transport rate offered by the porous layer of the undissolved mineral breaks the positive feedback between the fracture expansion and enhanced flow and solute transport, and hence suppresses the formation of wormhole in the initially preferred fracture. Besides, the total surface area increases as the dissolution proceeds, but mainly due to the exposure of the undissolved mineral to the pore space, as a result of the dissolution of the adjacent reactive mineral. The reactive surface area, however, reduces as the dissolution proceeds. The change trend of the reactive surface area of the porous medium studied in the present study is better calculated by the separated sphere model.

Simulations presented in this study are all two dimensional. Three dimensional pore-scale simulations of reactive transport processes in realistic rocks with multiple minerals are ongoing and will be presented in future publications.

## Acknowledgement

The authors acknowledge the support of LANL's LDRD Program, Institutional Computing Program, and National Nature Science Foundation of China (No. 51406145 and No. 51136004). W.Q. Tao thanks the support of National Basic Research Program of China (973 Program) (2013CB228304).The authors also thank Prof. L. Luquot from IDAEA, Geosciences, Barcelona, Spain for useful discussions.

# Appendix

Fig. A1 schematically shows fluid flow through a clear flow region sandwiched by two porous medium. In Fig. A1, the height of the clear flow region is $h$, and the height of both the two porous medium is $h_p$. For the convenience of discussion, $y=0$ is located at the center of the clear flow region, as shown in Fig. A1. A uniform pressure gradient is applied in $x$ direction in both the clear flow region and the porous medium. The porous medium is homogeneous and isotropic, and fluid flow in it is governed by Darcy's Law

$$u_p = -\frac{k_p}{\mu}\frac{dp}{dx} \tag{A1}$$

with $k_p$ as the permeability of the porous medium and $u_p$ as the superficial velocity in the porous medium. In the clear flow region, the governing equation of fluid flow is

$$\frac{d^2u}{dy^2} = \frac{1}{\mu}\frac{dp}{dx} \tag{A2}$$

At the interface between the clear flow region and the porous medium, additional boundary conditions must be applied to couple the flows in the two regions. In the present study, the semi-empirical slip boundary condition proposed by Beavers and Joseph [*Beavers and Joseph*, 1967 ] is employed

$$\frac{du}{dy}\bigg|_{y=\frac{h}{2}} = -\frac{\alpha}{\sqrt{k_p}}(u_b - u_p), \quad \frac{du}{dy}\bigg|_{y=-\frac{h}{2}} = \frac{\alpha}{\sqrt{k_p}}(u_b - u_p) \tag{A3}$$

with $u_b$ is the fluid flow at the interface. $\alpha$ is a dimensionless quantity depending on the structures of the porous medium. Therefore, the velocity in the clear flow region can be solved as

$$u = \frac{1}{2\mu}\frac{dp}{dx}y^2 - (\frac{1}{2\mu}\frac{\sqrt{k_p}}{\alpha}h + \frac{1}{8\mu}h^2 + \frac{k_p}{\mu})\frac{dp}{dx} \tag{A4}$$

The permeability of fluid flow through the clear flow region and the two porous medium, *k*, can be calculated by

$$k = -(\frac{Q+Q_p}{h+2h_p})\mu(\frac{dp}{dx})^{-1} \tag{A5}$$

where $Q$ and $Q_p$ is the flux in the clear region and porous medium, respectively. $Q$ is determined by

$$\begin{aligned} Q_1 &= \int_{-h/2}^{h/2} \left[ \frac{1}{2\mu}\frac{dp}{dx}y^2 - (\frac{1}{2\mu}\frac{\sqrt{k_p}}{\alpha}h + \frac{1}{8\mu}h^2 + \frac{k_p}{\mu})\frac{dp}{dx} \right] dy \\ &= \frac{1}{6\mu}\frac{dp}{dx}\frac{h^3}{4} - (\frac{1}{2\mu}\frac{dp}{dx}\frac{\sqrt{k_p}}{\alpha}h^2 + \frac{1}{8\mu}\frac{dp}{dx}h^3 + \frac{k_p}{\mu}\frac{dp}{dx}h) \end{aligned} \tag{A6}$$

and $Q_p$ is calculated by

$$Q_p = 2u_p h_p \tag{A7}$$

Therefore, the permeability of the entire domain is

$$k = \frac{1}{h+2h_p}(\frac{h^3}{12} + \frac{1}{2}\frac{\sqrt{k_p}}{\alpha}h^2 + k_p h + 2k_p h_p) \tag{A8}$$

# Figure and Table Caption

Fig. 1 Eight directions for mineral growth in the QSGS reconstruction method or the D2Q9 lattice model in LBM

Fig. 2 Different morphologies of β in the rocks using the QSGS method. The black area represents mineral β, the gray part is mineral α and the white space is void space. The rocks have a fracture at the center of the domain. (a) Random β distribution: $c_d$=0.05, $P_{(1,2,3,4)}$= 0.0004, $P_{(5,6,7,8)}$= 0.0001, $\varepsilon_\beta$=0.2; (b) $x$-oriented β distribution: $c_d$=0.05, $P_{(1,3)}$= 0.0004, $P_{(2,4,5,6,7,8)}$= 0., $\varepsilon_\beta$=0.2; (c) $y$-oriented β distribution: $c_d$=0.05, $P_{(2,4)}$= 0.0004, $P_{(1,3,5,6,7,8)}$= 0., $\varepsilon_\beta$=0.2

Fig. 3 Temporal evolutions of the concentration distributions of $R_{(aq)}$ as well as the geometries of the rocks for $Pe$=0.011 and $Da$=0.01. In this scenario, the process is reaction-controlled, indicating the diffusion is much faster than the reaction. Fairly uniform distributions of $R_{(aq)}$ in the entire void space and thus uniform dissolution of mineral α can be observed. (a) t = 1000s, (b) t = 5000s, (c) t = 13000s

Fig. 4 Temporal evolutions of the concentration distributions of $R_{(aq)}$ as well as the geometries of the rocks for $Pe$=0.011 and $Da$=5. In this scenario, the process is diffusion-controlled, indicating the diffusion is slow compared to reaction. Dissolution mainly takes place at the α-fluid interface facing the inlet boundary. The dissolution front gradually advances towards the downstream, behind which the fracture-rock surface is intact. (a) t = 100s, (b) t = 1000s, (c) t = 6000s

Fig. 5 Temporal evolutions of the concentration distributions of $R_{(aq)}$ as well as the geometries of the rocks for $Pe$=5.5, and $Da$=5. In this scenario, the process is diffusion-controlled with fast flow rate. The dominated mechanism of mass transport is convection. The dissolution advances along the fracture-rock interface. (a) t = 50s, (b) t = 100s, (c) t = 250s

Fig. 6 Relationship between porosity and permeability for different $Pe$ and $Da$. (a) Rocks with mineral α and β with random β distribution, (b) Rocks with mineral α only

Fig. 7 Apertures. (a) The distribution of mechanical aperture along $x$ direction, (b) Comparison between changes in hydraulic, mechanical and chemical aperture

Fig.8 Relationship between porosity and permeability for un-dissolved mineral with different structures

Fig.9 Relationship between porosity and permeability for un-dissolved mineral with different dissolution rate

Fig. 10 Rocks with complex porous structures. The porous geometry consists of pieces of rocks of varying sizes surrounded by randomly oriented and interconnected micro fractures. Mineral β is randomly distributed inside the rocks with a volume fraction of 0.3.

Fig. 11 Temporal evolutions of the concentration distributions of $R_{(aq)}$ as well as the geometries of the rocks for $Pe$=6.011 and $Da$=5.0. Fingering-like contours with a sharp front in the fractures clearly show convection as the dominant mechanism of local mass transport ($t$ = 2s and $t$=200s). Positive feedback between the fluid flow and rock dissolution is not observed. The "wormholing" phenomenon is somewhat suppressed when undissolved minerals exists. (a) $t$ = 2s, (b) $t$ = 200s, (c) $t$ = 500s, (d) $t$ = 1000s, (d) $t$ = 2000s, (e) $t$ = 5600s

Fig. 12 Temporal evolutions of the porosity and permeability

Fig. 13 Variation of the surface area with time

Fig. A1 Schematic of fluid flow through free region sandwiched by porous medium

Table 1 Parameter values used in the simulation

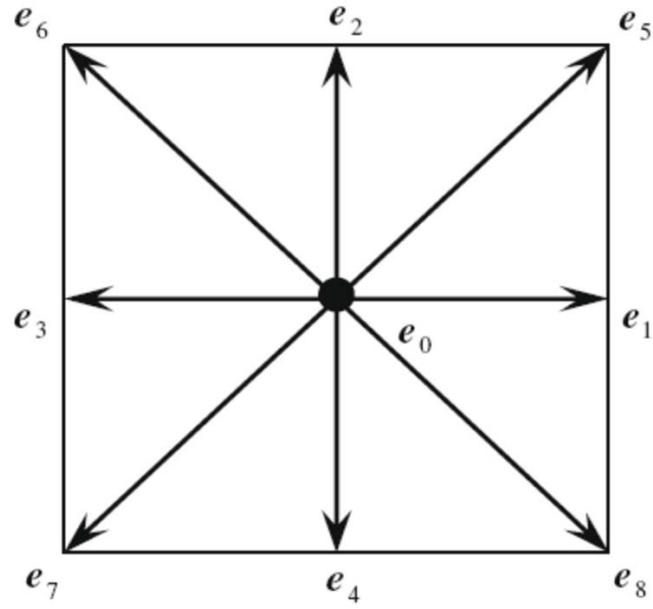

Fig. 1 Eight directions for mineral growth in the QSGS reconstruction method or the D2Q9 lattice model in LBM

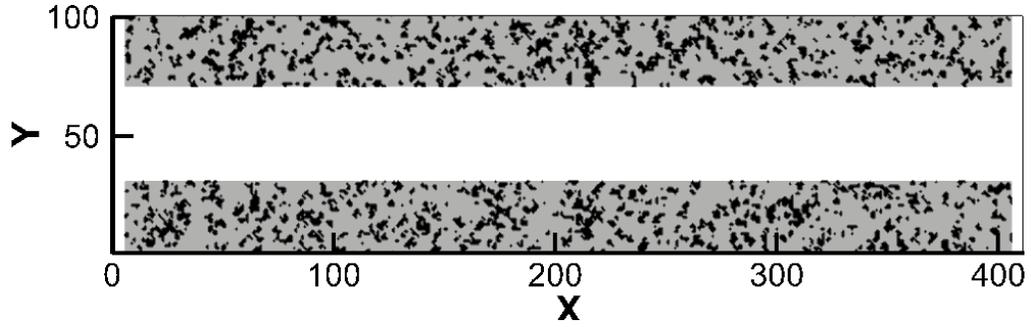

(a) Random β distribution: $c_d=0.05$, $P_{(1,2,3,4)}= 0.0004$, $P_{(5,6,7,8)}= 0.0001$, $\varepsilon_\beta=0.2$

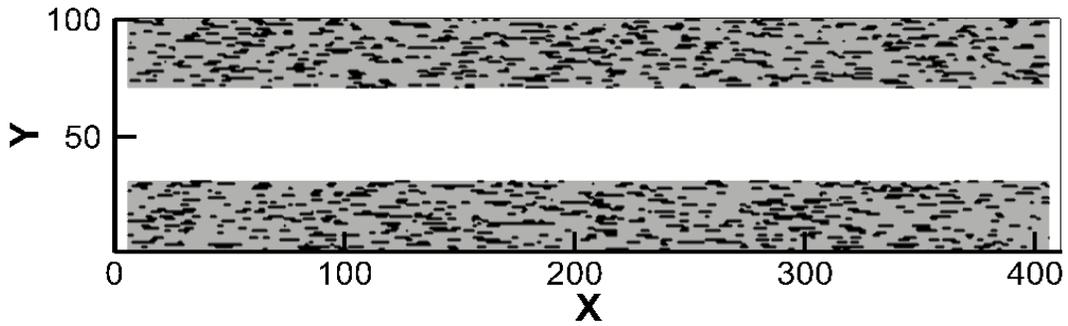

(b) x-oriented β distribution: $c_d=0.05$, $P_{(1,3)}= 0.0004$, $P_{(2,4,5,6,7,8)}= 0.$, $\varepsilon_\beta=0.2$

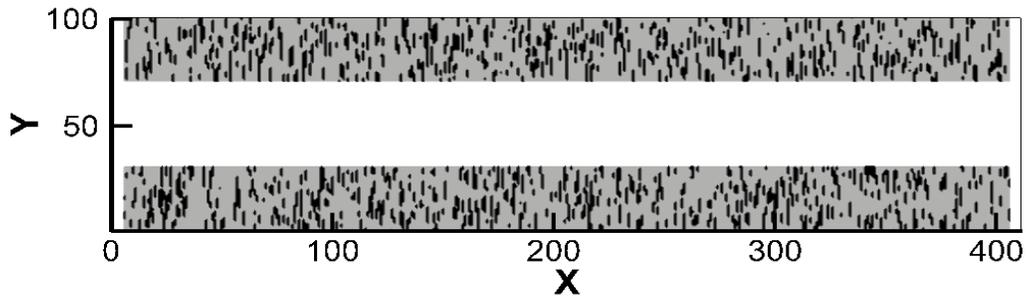

(c) y-oriented β distribution: $c_d=0.05$, $P_{(2,4)}= 0.0004$, $P_{(1,3,5,6,7,8)}= 0.$, $\varepsilon_\beta=0.2$

Fig. 2 Different morphologies of β in the rocks using the QSGS method. The black area represents mineral β, the gray part is mineral α and the white space is void space. The rocks have a fracture at the center of the domain

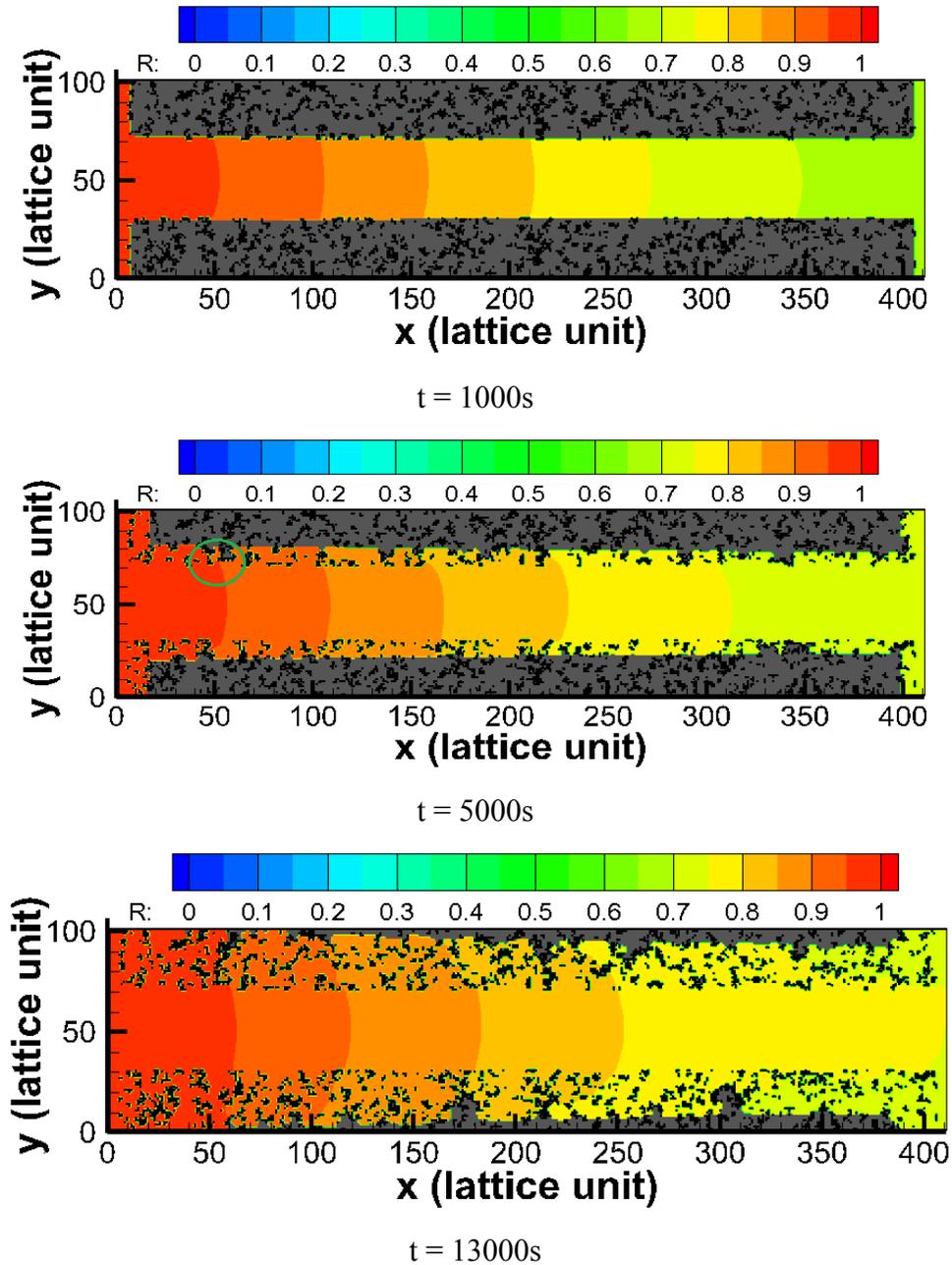

Fig. 3 Temporal evolutions of the concentration distributions of $R_{(aq)}$ as well as the geometries of the rocks for $Pe$=0.011 and $Da$=0.01. In this scenario, the process is reaction-controlled, indicating the diffusion is much faster than the reaction. Very uniform distributions of $R_{(aq)}$ in the entire void space and thus uniform dissolution of mineral α can be observed.

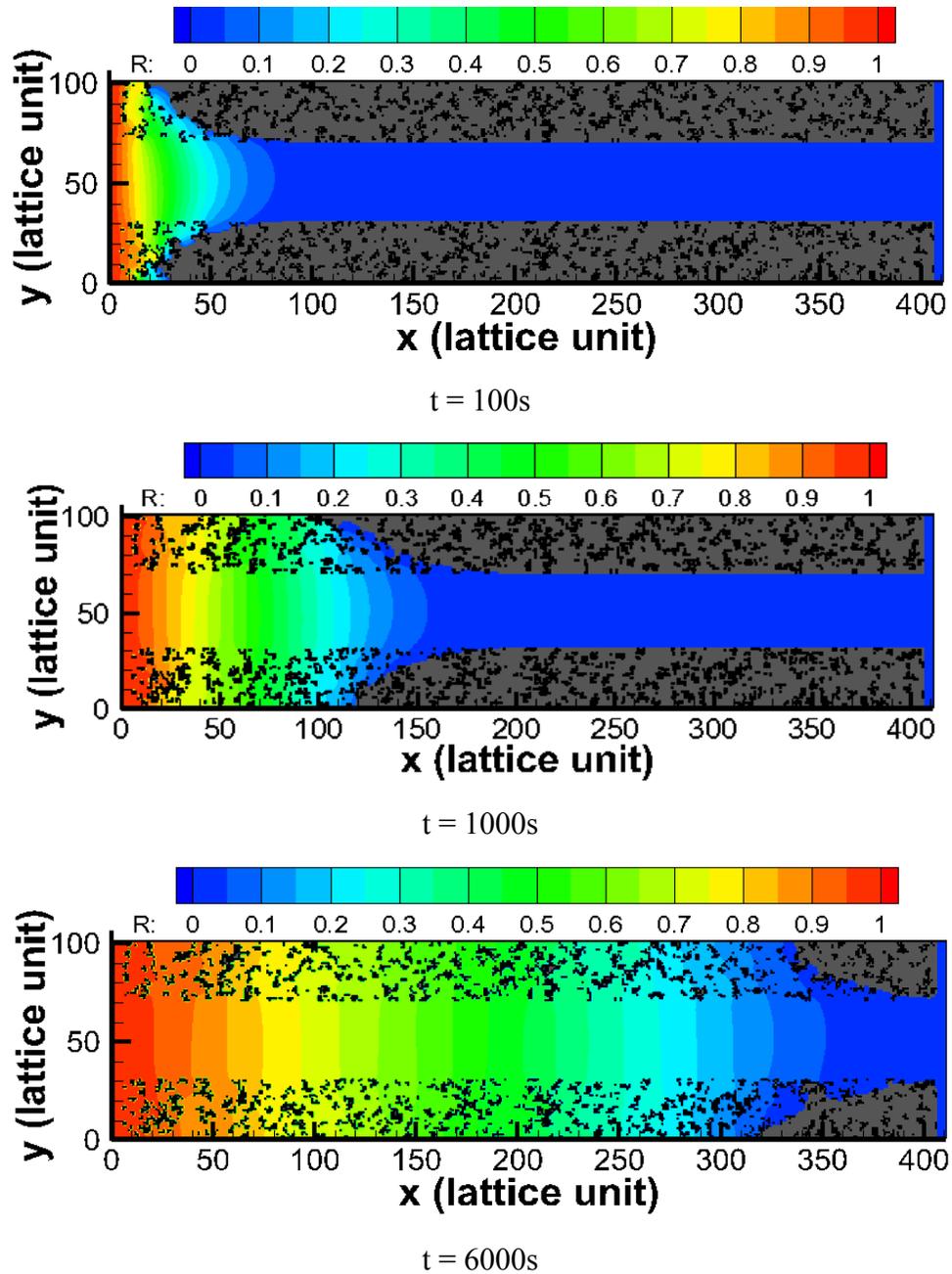

Fig. 4 Temporal evolutions of the concentration distributions of $R_{(aq)}$ as well as the geometries of the rocks for $Pe$=0.011 and $Da$=5. In this scenario, the process is diffusion-controlled, indicating the diffusion is slow compared to reaction. Dissolution mainly takes place at the α-fluid interface facing the inlet boundary. The dissolution front gradually advances towards the downstream (Fig. 4(b-c)), behind which the fracture-rock surface is intact.

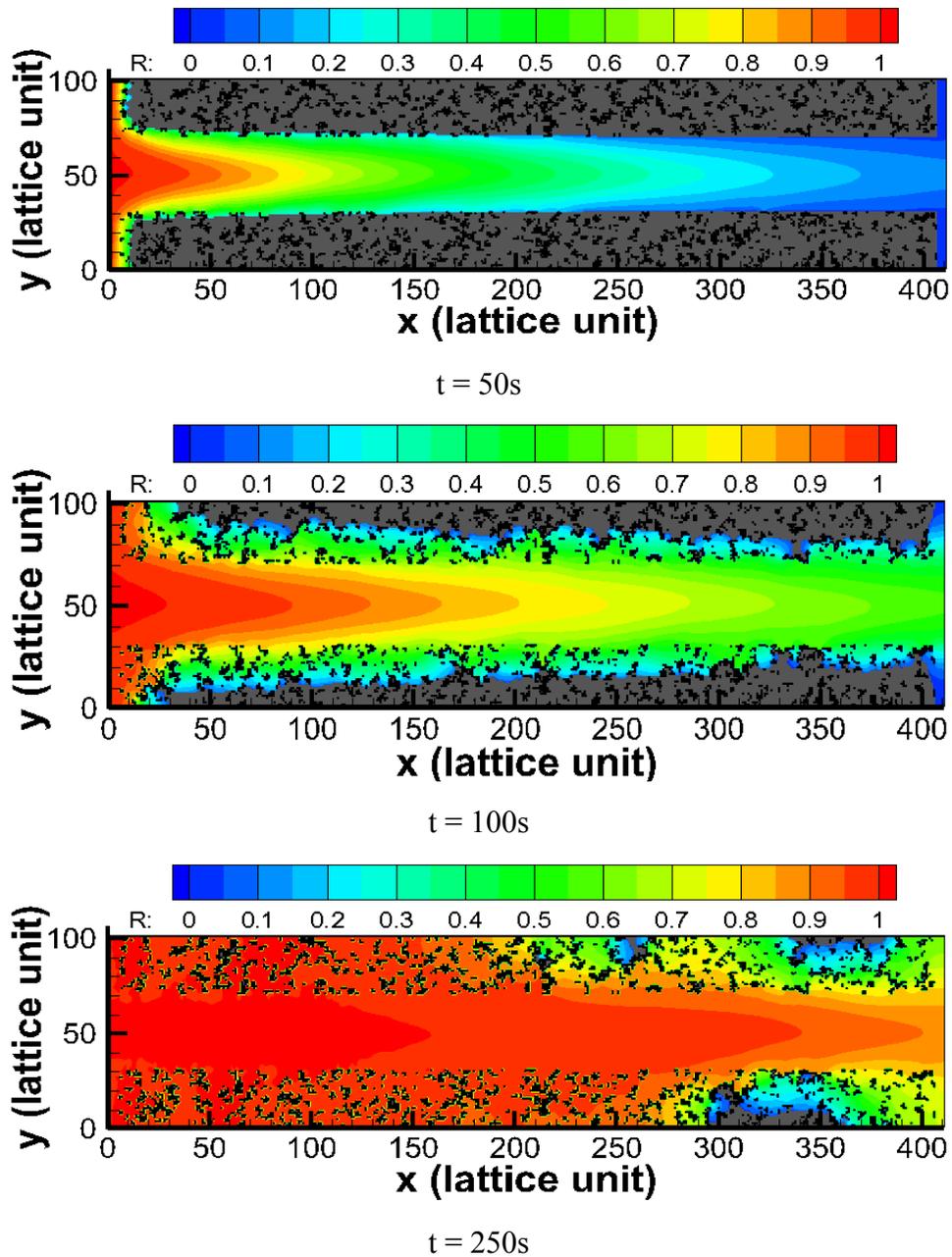

Fig. 5 Temporal evolutions of the concentration distributions of R$_{(aq)}$ as well as the geometries of the rocks for *Pe*=5.5, and *Da*=5. In this scenario, the process is diffusion-controlled with fast flow rate. The dominated mechanism of mass transport is convection. The dissolution advances along the fracture-rock interface

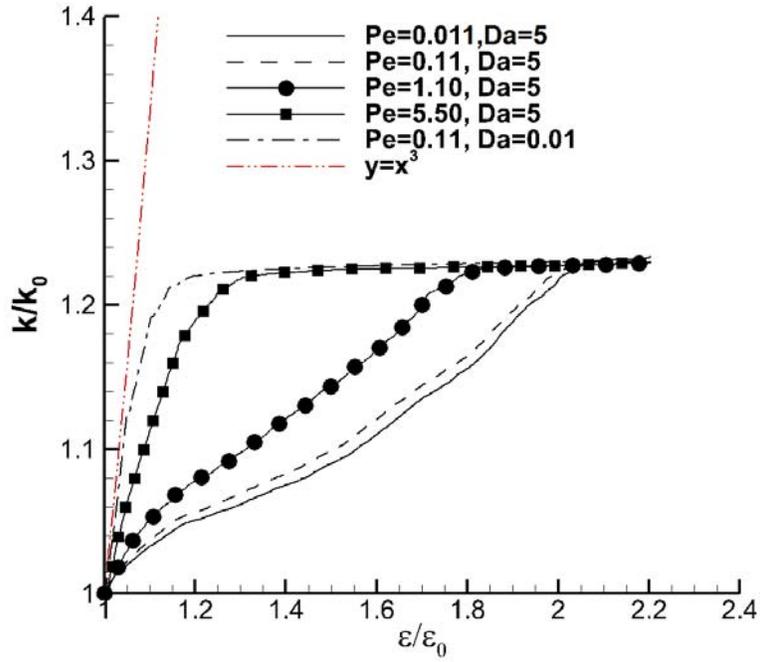

(a) Rocks with mineral α and β for the case with random β distribution

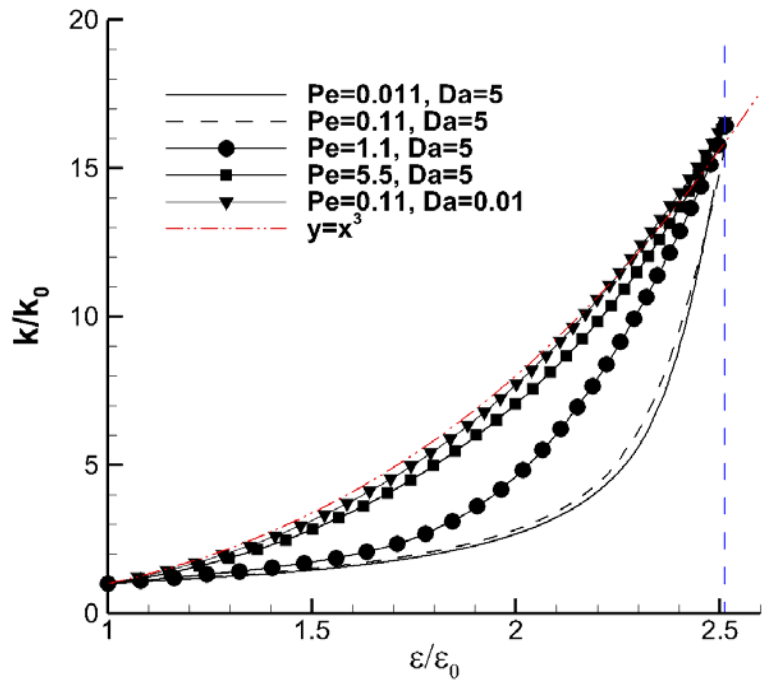

(b) Rocks with mineral α only

Fig. 6 Relationship between porosity and permeability for different *Pe* and *Da*

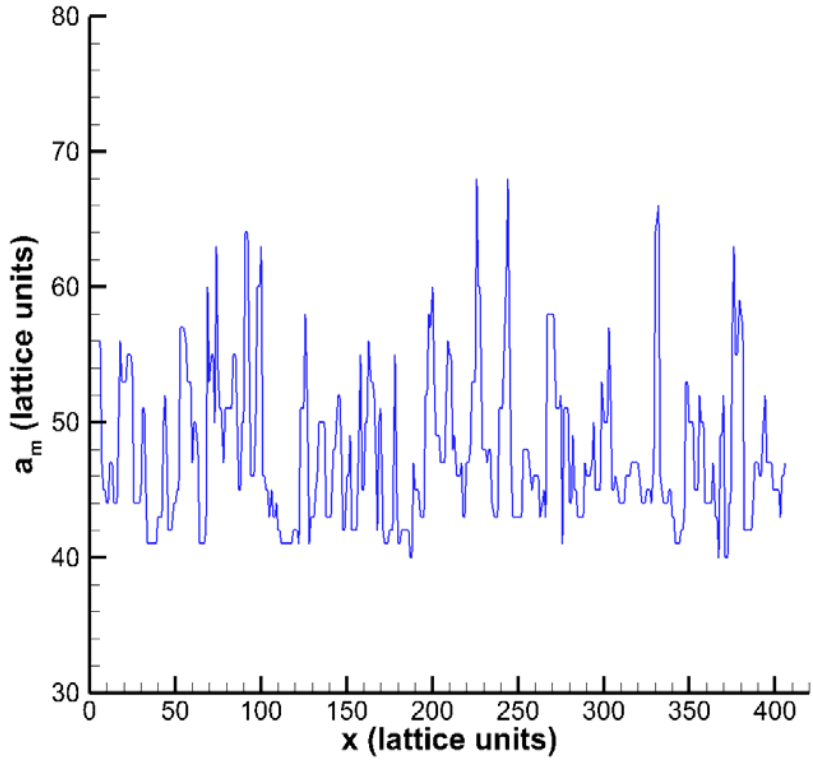

(a) The distribution of mechanical aperture along *x* direction

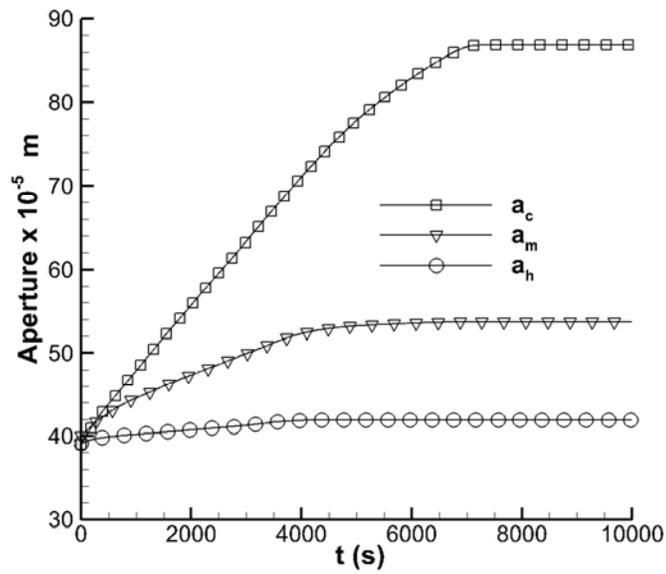

(b) Comparison between changes in hydraulic, mechanical and chemical aperture

Fig. 7 Apertures

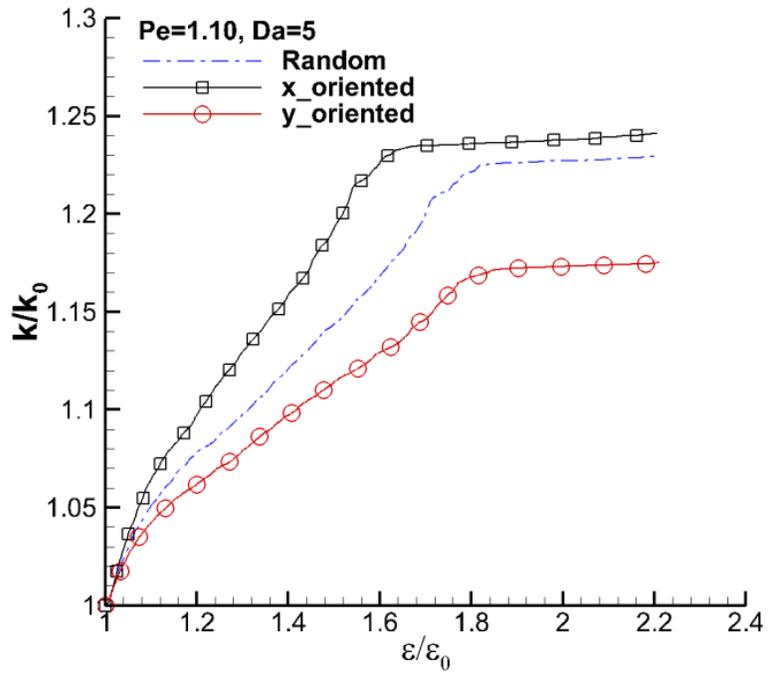

Fig.8 Relationship between porosity and permeability for un-dissolved mineral with different structures

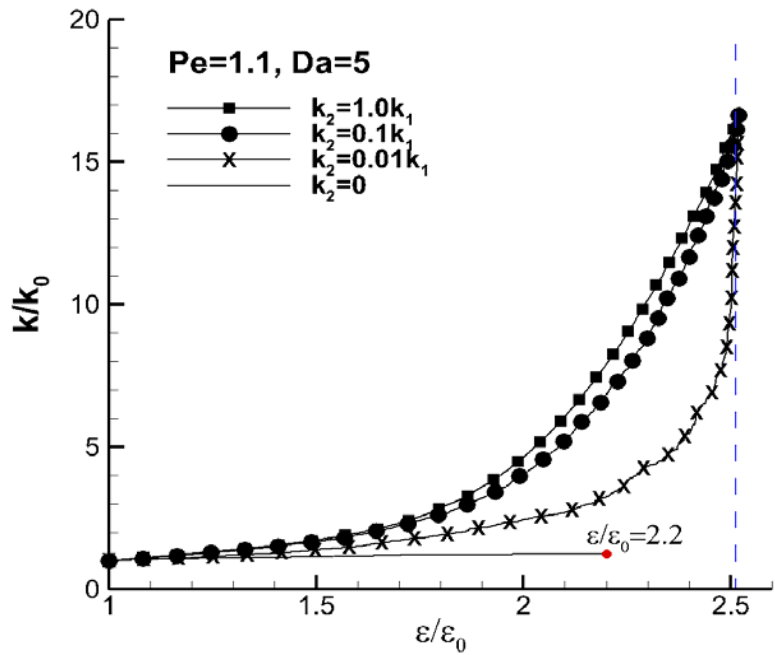

Fig.9 Relationship between porosity and permeability for un-dissolved mineral with different dissolution rate

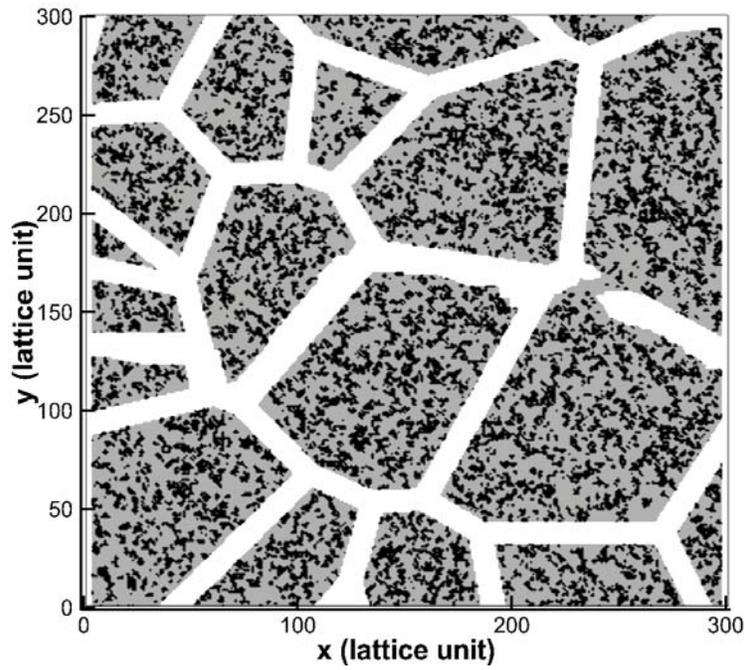

Fig. 10 Rocks with complex porous structures. The porous geometry consists of pieces of rocks of varying sizes surrounded by randomly oriented and interconnected micro fractures. Mineral β is randomly distributed inside the rocks with volume fraction of 0.3.

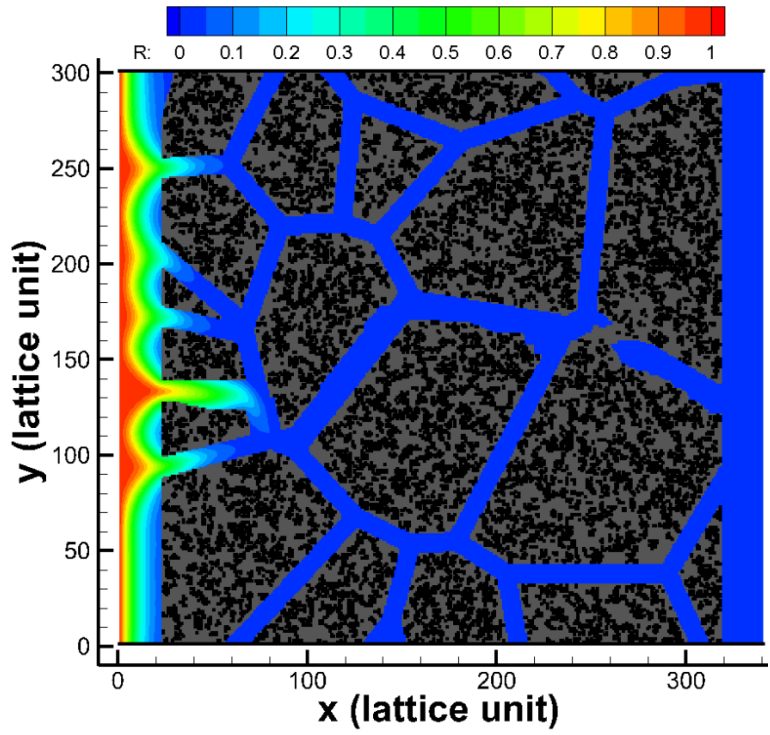

*t* = 2s

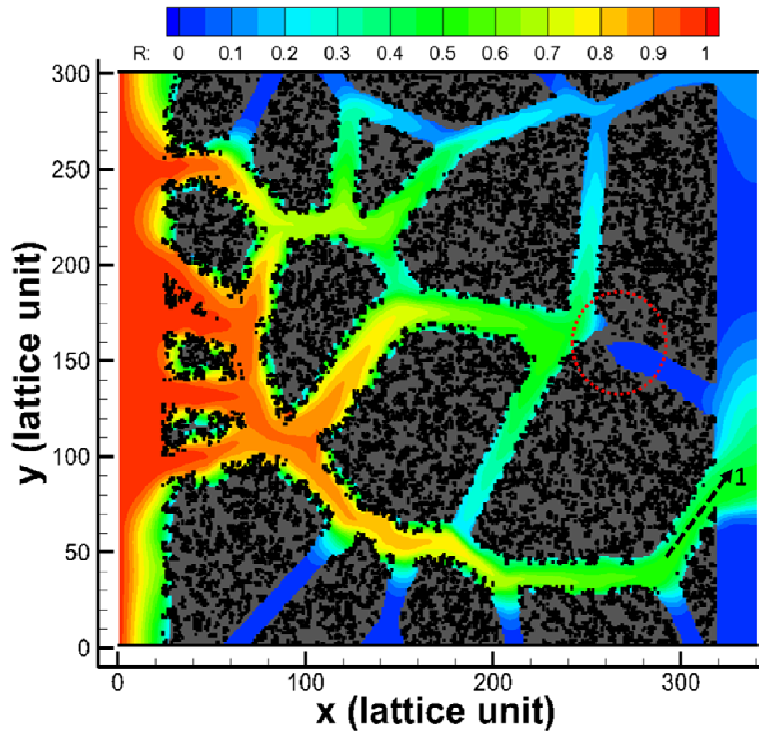

*t* = 200s

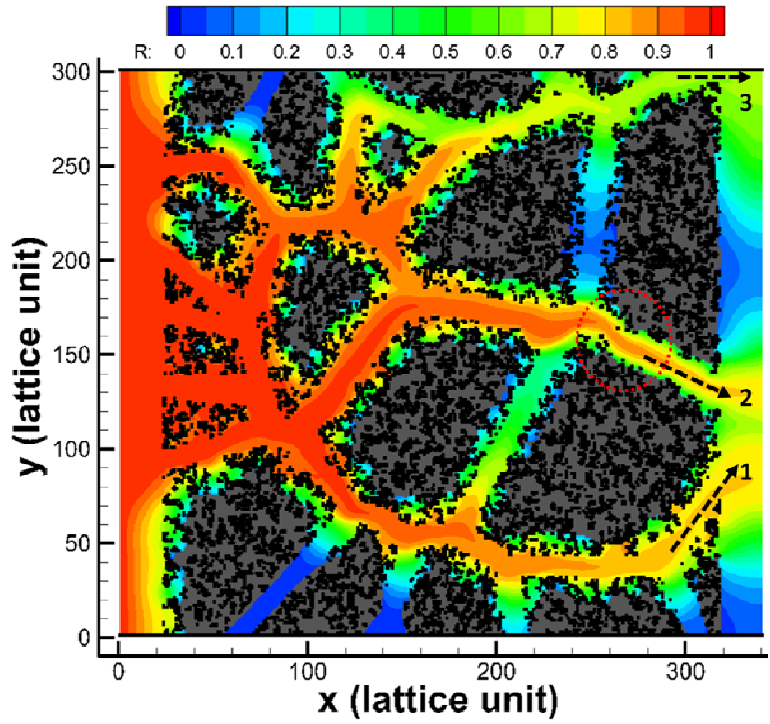

$t = 500\text{s}$

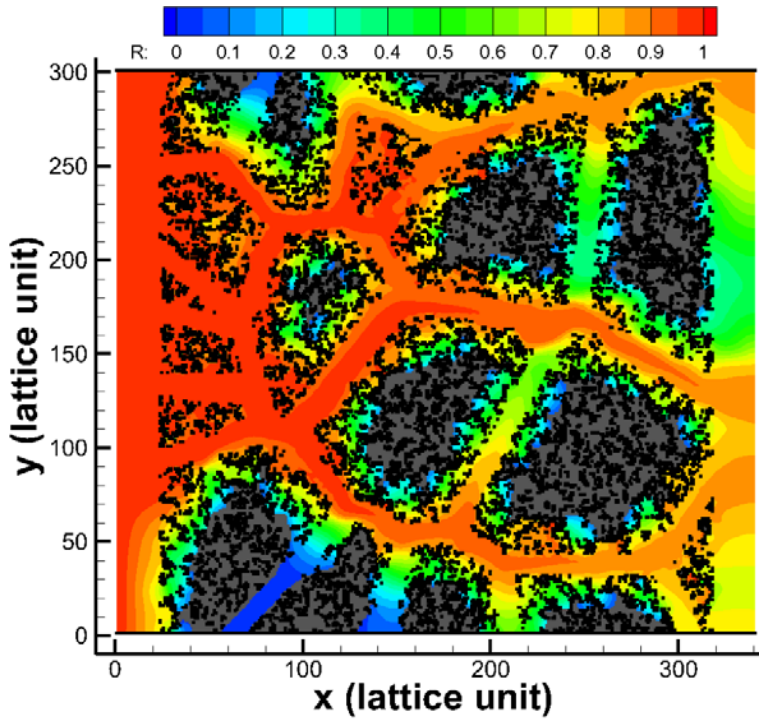

$t = 1000\text{s}$

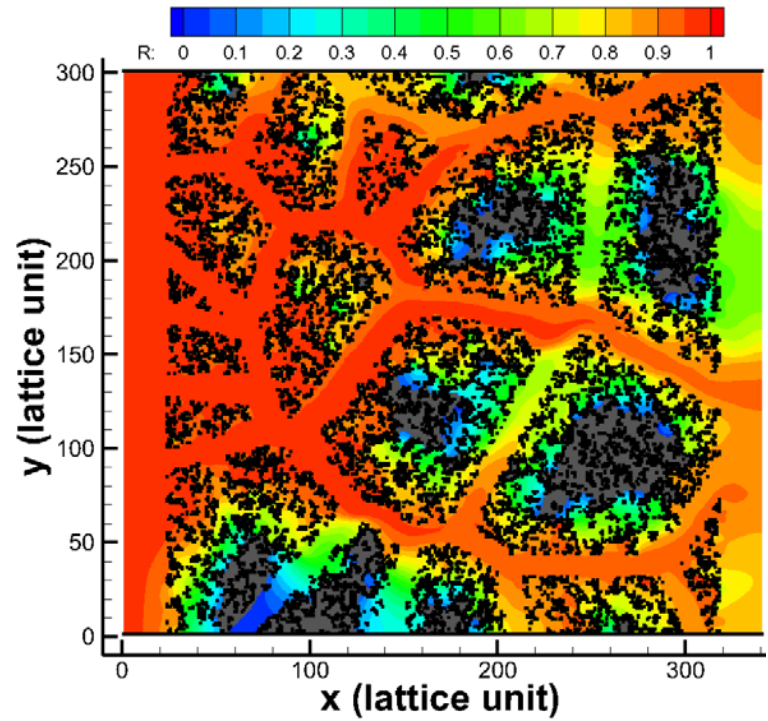

*t* = 2000s

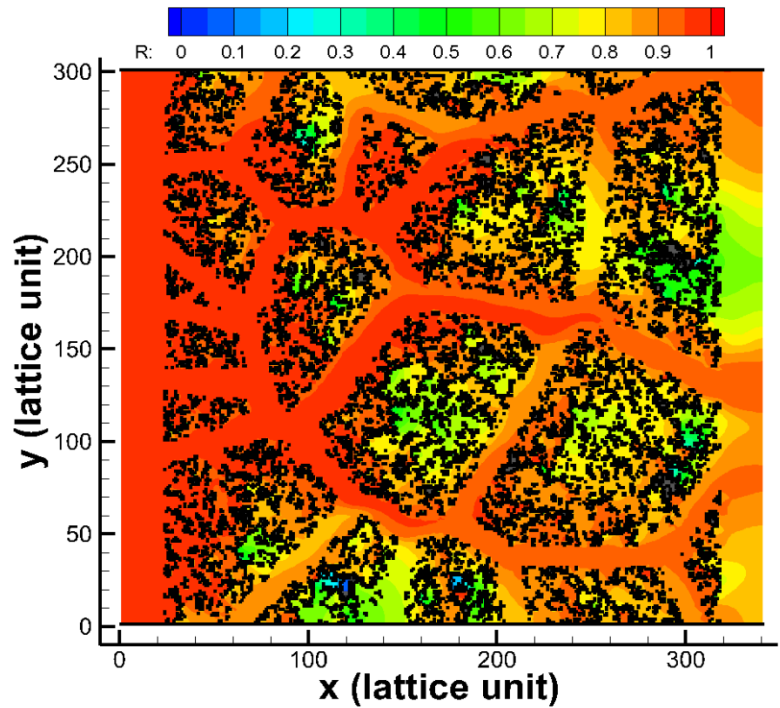

*t* = 5600s

Fig. 11 Temporal evolutions of the concentration distributions of $R_{(aq)}$ as well as the geometries of the rocks for *Pe*=6.011 and *Da*=5.0. Fingering-like contours with a sharp front in the fractures

clearly prove convection as the dominant mechanism of local mass transport ($t = 2$s and $t=200$s). Positive feedback between the fluid flow and rock dissolution is not observed. The "wormholing" phenomenon is somewhat suppressed when undissolved minerals exists.

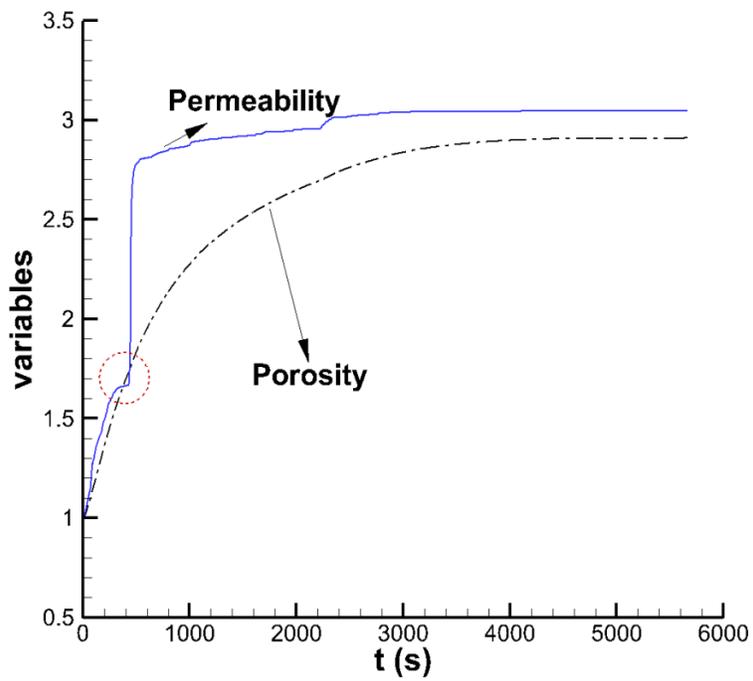

Fig. 12 Temporal evolutions of the porosity and permeability

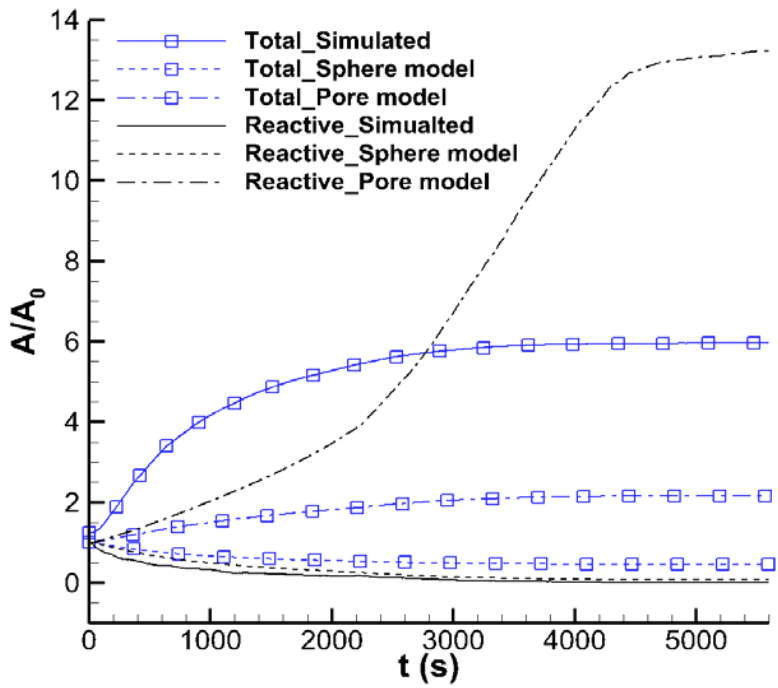

Fig. 13 Variation of the surface area with time

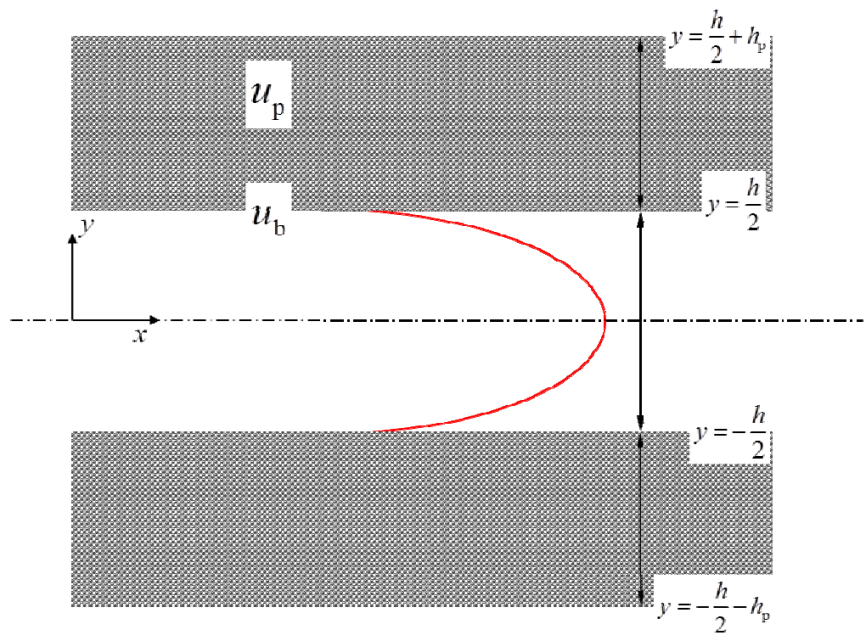

Fig. A1 Schematic of fluid flow through free region sandwiched by porous medium

Table 1 Parameter values used in the simulation

| Physicochemical variables | | Value of Physical unit | Value of lattice unit |
|---|---|---|---|
| Kinematics viscosity | $\upsilon$ | $1.0 \times 10^{-6}$ m$^2$ s$^{-1}$ | 1/6 |
| Diffusivity of R$_{(aq)}$ and P$_{(aq)}$ | $D$ | $2.0 \times 10^{-9}$ m$^2$ s$^{-1}$ | Depending on simulation cases [a] |
| Molar volume of mineral α | $M$ | $3.75 \times 10^{-5}$ m$^3$ mol$^{-1}$ | $3.75 \times 10^{-2}$ |
| Inlet concentration of R$_{(aq)}$ | $C_{R,in}$ | 1M (1000 mol m$^{-3}$) | 1 |
| Inlet concentration of P$_{(aq)}$ | $C_{P,in}$ | 0M (0 mol m$^{-3}$) | 0 |
| Effective reaction rate constant | $k_r$ | Depending on simulation cases, $k_r = 5 \times 10^{-6} Da$ m s$^{-1}$ | |
| Averaged fluid flow velocity | $\bar{u}$ | Depending on simulation cases, $\bar{u} = 5 \times 10^{-6} Pe$ m s$^{-1}$ | |
| Effective equilibrium constant | $K_{eq}$ | $1 \times 10^{10}$ | $1 \times 10^{10}$ |
| **Variables in LB models** | | | |
| Collision time in Eq.(6) | $\tau_\nu$ | 1 | |
| $\sigma$, $\lambda$ and $\gamma$ in eq. (8) | | 5/12, 1/3, 1/12 | |
| Collision time in Eq. (13) | $\tau_{g,\alpha}$, $\tau_{g,\beta}$ | Depending on simulation cases | |
| Rest fraction | $J_{0,\alpha}$, $J_{0,\beta}$ | 0.2, 0.2 | |

[a] Depending on simulation cases means the value varies for different simulation cases for the purpose to obtain desirable $Pe$ and $Da$.